\documentclass[10pt,journal, twoside]{IEEEtran}
\usepackage[T1]{fontenc}
\usepackage{multirow}
\usepackage{mathtools}
\usepackage{commath}
\usepackage{color}
\usepackage{xcolor}
\usepackage{comment}
\usepackage{subfigure}
\usepackage{algorithm}
\usepackage{algcompatible}
\usepackage{algpseudocode}
\usepackage{float}
\usepackage{array}
\newcolumntype{P}[1]{>{\centering\arraybackslash}p{#1}}

\usepackage{amsmath,amssymb,amsfonts}

\usepackage{graphicx}
\usepackage{textcomp}

\usepackage[numbers]{natbib}
\usepackage{enumitem}
\setlist[itemize]{leftmargin=*}
\def\BibTeX{{\rm B\kern-.05em{\sc i\kern-.025em b}\kern-.08em
    T\kern-.1667em\lower.7ex\hbox{E}\kern-.125emX}}

\newcommand{\comg}{\textcolor{black}} %green 

\begin{document}
%\history{Date of publication xxxx 00, 0000, date of current package xxxx 00, 0000.}
%\doi{10.1109/ACCESS.2019.DOI}

\title{Vulnerability-Aware Resilient Networks: \\
Software Diversity-based Network Adaptation}
\author{Qisheng Zhang, Jin-Hee Cho, \IEEEmembership{Senior Member, IEEE,} Terrence J. Moore, \IEEEmembership{Member, IEEE},  Ing-Ray Chen, \IEEEmembership{Member, IEEE}
\IEEEcompsocitemizethanks{\IEEEcompsocthanksitem
Qisheng Zhang, Jin-Hee Cho, and Ing-Ray Chen are with the Department of Computer Science, Virginia Tech, Falls Church, VA, USA. Email: \{qishengz19, jicho, irchen\}@vt.edu.  Terrence J. Moore is with US Army Research Laboratory, Adelphi, MD, USA. Email: terrence.j.moore.civ@mail.mil.}
}
%\tfootnote{add funding information later}

\markboth{IEEE Transactions on Network Service and Management, vol. X, no. X, Month, 2020}{Zhang \MakeLowercase{\textit{et al.}}: Vulnerability-Aware Resilient Networks:
Software Diversity-based Network Adaptation}
\maketitle
%\IEEEtitleabstractindextext{
\begin{abstract}
By leveraging the principle of software polyculture to ensure security in a network, we proposed a vulnerability-based software diversity metric to determine how a network topology can be adapted to minimize security vulnerability while maintaining maximum network connectivity.  Our proposed software diversity-based adaptation (SDA) scheme estimates a node's software diversity based on the vulnerabilities of software packages installed on other nodes on attack paths reachable to the node and employs it for edge adaptations, such as removing an edge with a neighboring node that exposes high security vulnerability because two connected nodes use the same software packages or a neighboring node may have high software vulnerability or adding an edge with another node with less or no security vulnerability because the two nodes use different software packages or have low vulnerabilities associated with them.  To validate the proposed SDA scheme, we conducted extensive experiments comparing the proposed SDA scheme with counterpart baseline schemes in real networks.  Our simulation experimental results proved the outperformance of our proposed SDA compared to the existing counterparts and provided insightful findings in terms of the effectiveness and efficiency of the proposed SDA scheme under three real network topologies with vastly different network density. 
\end{abstract}
%}

%\maketitle

%\IEEEdisplaynontitleabstractindextext
\IEEEpeerreviewmaketitle

\section{Introduction}
\label{sec:introduction}

\subsection{Motivation}

Inspired by the close relationship between the diversity of species and the resilience of ecosystems~\cite{Walker99}, information and software assurance research has evolved to include the concept of {\em software diversity} for enhanced security~\cite{Hole15, Hole15-antifragile, Larsen14, Larsen15, Donnell04}. Due to the dominant trend of software monoculture deployment for efficiency and effectiveness of service provisions, attackers have been granted significant advantages in that acquiring the intelligence needed to compromise a single software vulnerability enables the capability of efficiently compromising other homogeneous system components, such as operating systems, software packages, and/or hardware packages~\cite{Yang16}.  To deny this advantage, the concept of diversity has been applied in the cybersecurity literature~\cite{Knight16}. Randomization of software features has been used to thwart cyber attacks by increasing uncertainty towards a target system whose critical information was known to an attacker previously. The concept of moving target defense (MTD)~\cite{Hong16, Manadhata11} has been proposed to change the attack surface in order to increase uncertainty and confusion for attackers and software diversity-based security mechanisms have also been used as part of MTD techniques. 

Research has shown that software diversity is closely related to enhancing the immunization of a computer system that halts multiple outbreaks of malware infections simultaneously occurring with heterogeneous and sparse spreading patterns~\cite{Newman10}. Hence, the rationale that software diversity reduces malware spreading is quite well known and has been validated for its effectiveness to some extent~\cite{Hole15, Hole15-antifragile}.  This underlying philosophy encompasses a simple principle: {\em software polyculture} enhances security~\cite{Hole15}. Due to the accessibility to the  Internet, which enables the distribution of individualized software and cloud computing with the computational power to perform diversification, massive-scale software diversity is becoming a realistic and practical approach to enhance security~\cite{Larsen14}. Although the benefit of software diversity seems obvious, its secure and transparent implementation of automatic software diversity is highly challenging~\cite{Larsen15}. In addition, no prior work has considered software diversity metrics as the basis to adapt a network topology to balance network connectivity and system security where each node's software vulnerability is incorporated into estimating each node's software diversity.

In this work, we are interested in developing a software diversity metric to measure a node's software diversity based on software vulnerabilities of intermediate nodes on attack paths reachable to the node. 
\vspace{-2mm}
\subsection{Research Problem}
In this work, we develop a software diversity metric for measuring a network topology in terms of minimizing security vulnerabilities against epidemic attacks (e.g., malware/virus spreading) while maintaining a sufficient level of network connectivity to provide seamless service availability. The proposed software diversity metric can be used to make decisions related to which two nodes should be disconnected or connected in order to construct an improved network topology meeting these two goals, minimizing security vulnerability and maximizing network connectivity.  However, identifying the optimal network topology requires an exponential solution complexity~\cite{Yang08}. In this work, we propose a heuristic method called software diversity-based adaptation (SDA) to generate a better network topology that is resilient against epidemic attacks with a sufficiently high network connectivity where the deployment cost is acceptable.    
We leverage percolation theory~\cite{Newman10}, which has been used to describe the process or paths of some liquid passing through a medium.  We use this theory to model and analyze attack processes and defense or recovery processes by using site or bond percolation.  {\em Site percolation} (i.e., removing a node)~\cite{Newman10} is used to model an attacker's behavior in compromising another node, implying that the node being percolated refers to the node being compromised (infected) by the attacker, leading to the disconnection of all edges around the node to reflect its failure or its being detected by an intrusion detection system (IDS). {\em Bond percolation} is used to adjust edges between nodes such that connected nodes with high security vulnerability (e.g., two connected nodes have the same software package installed or a neighbor node has high software vulnerability) are disconnected while disconnected nodes with low or no security vulnerability (e.g., two disconnected nodes using a different software with low software vulnerability) can be connected in a given network.

\begin{comment}
\subsubsection{Research Questions} We aim to answer the following {\bf key research questions}:
\begin{itemize}
\item Is our proposed software diversity metric a good indicator to represent network resilience in both security and network connectivity in the presence of epidemic attacks? 
\item What are the key impacts from varying the key design parameters in terms of the attack density (or strength), the number of software packages available, the software diversity level, or the network adaptation deployment cost when the proposed scheme is compared against comparable counterpart baseline schemes? 
\item Which adaptation strategies perform the best among all the schemes considered and under what circumstances (e.g., network density, attack density, or availability of software packages)? 
\item What are the key factors that maximize network resilience in security and network connectivity while incurring acceptable deployment cost? 
\end{itemize}
\end{comment}

\vspace{-2mm}
\subsection{Key Contributions}

We made the following {\bf key contributions} in this work:
\begin{itemize}
\item This work is the first that takes a multidisciplinary approach by considering both the computer science's software diversity to enhance cybersecurity and percolation theoretic network resilience techniques to study the effect of interconnectivity on network connectivity under epidemic attacks. To be specific, we develop network adaptation strategies that determine whether to add or remove edges between two nodes in a given network, aiming to minimize network vulnerabilities against epidemic attacks while maintaining maximum network connectivity.  Given that each node is installed with a set of software (we call it a `software package'), we investigate network resilience and vulnerability depending on how a network topology is connected under epidemic attackers who can exploit the vulnerabilities based on their knowledge on software vulnerabilities.  

\item We develop a novel software diversity metric that measures a node's software diversity level, representing both the vulnerabilities of attack paths reachable to the node and the network connectivity.  To minimize computational complexity in estimating the node's software diversity based on attack path vulnerability, we introduce a node's local network only based on the node's $k$-hop neighbors.  This approach allows us to provide a lightweight method to compute each node's software diversity.  To prove the effectiveness of this software diversity metric, we use it as the criterion to determine whether to add/remove an edge between two nodes. 
\item Although most software diversity-based network topology adaptations are studied by shuffling the types of software packages~\cite{Yang08, Yang16}, our work takes one step further by changing network topology, which is proven much more effective than its software shuffling counterpart (e.g., graph coloring) in reducing vulnerability to epidemic attacks while maximizing the network connectivity.  In addition, our proposed software diversity-based network adaptations are lightweight showing acceptable operational cost while achieving minimum security vulnerability and maximum network connectivity, which opens a door for the applicability in resource-constrained, contested network environments.
\item We validate the outperformance of the proposed SDA strategy by conducting a comprehensive comparative performance analysis with the following six schemes (see Section \ref{subsec:comparing_schemes}): non-adaptation, random adaptation, graph-coloring, and three variants of the proposed SDA strategies.  We analyze the effect of key design parameters such as network density, attack density, and the number of software packages available on four performance metrics (see Section~\ref{subsec:metrics}), i.e., the size of a giant component, the fraction of undetected compromised nodes, software diversity levels, and defense cost (i.e., shuffling plus network topology adaptation costs).  We validate the outperformance of our SDA scheme in three real network topologies covering dense (high), medium dense, and sparse (low) networks~\cite{snapnets}.  Further, to profoundly understand the effect of various network characteristics, we conduct sensitivity analysis under a random graph using the Erd{\"o}s-R{\'e}nyi (ER) network model and analyze the results.  Due to space constraints, we placed these results for the ER network in Sections C.2--C.3 of the appendix file. %\tjm{[<-Are the prior three key contributions a little long?]}
\end{itemize}
We will discuss the answers of the research questions in Section~\ref{sec:exp-result-analysis} and conclude them in Section~\ref{sec:conclusion}. %\tjm{[<-This can be stated in an outline of the work that might be placed at the end of the introduction.]}

%\tjm{[Should there be an outline to the paper here?]}
%\subsection{Paper Structure}
%The rest of this paper is organized as follows. Section \ref{sec:related_work} discusses the overview of the state-of-the-art related work in terms of percolation theory and software diversity as one of MTD strategies. Section \ref{sec:system_model} describes the proposed system model including the models of a network, a node, an attack, and a defense. Section \ref{sec:software_diversity_strategies} provides the detail of the proposed software diversity metric and the proposed SDA strategy. Section \ref{sec:results} demonstrates the experimental results comparing the performance of the six schemes including the proposed SDA schemes and other baseline and counterparts. Section \ref{sec:conclusion} concludes our paper and suggests our future work directions.

\section{Background \& Related Work} \label{sec:related_work}
This section provides an overview of related work and background literature in terms of the percolation theory studied for network resilience in Network Science and the software diversity-based approaches studied for system security in Computer Science.

\subsection{Percolation Theoretic Network Resilience} \label{subsec:percolation_theory}
Percolation theory has been substantially used to investigate network resilience (or robustness) in Network Science. {\em Site percolation} and {\em bond percolation} are commonly used to select a node or an edge to remove or add, to model the choice of nodes to immunize in the context of epidemics on networks, such as disease transmission, computer malware/virus spreading, or behavior propagation (e.g., product adoption)~\cite{Dezso02, Newman10}. Recently, percolation theory was leveraged to develop software diversity techniques particularly to solve a software assignment problem~\cite{Yang08, Yang16} because how nodes are connected matters in propagating malware infection while choosing nodes or edges to add or remove is exactly following the concept of site or bond percolation in percolation theory~\cite{Newman10}. 

The origins of percolation theory come from the mathematical formalization of statistical physics research on the flow of liquid through a medium~\cite{Grimmett1997}. Percolation theory has been substantially applied to networks to study connectivity, robustness~\cite{Barabasi2016, Newman10}, reliability~\cite{Li15}, and epidemics~\cite{cardy1985epidemic, moore2000epidemics}. The percolation process was studied in computer science under the notion of ``network resilience''~\cite{Colbourn87, Najjar90, Sterbenz10}, independent of its development in the statistical physics literature. More recent developments in the physics literature have profoundly influenced studies in the computer science. These contributions have incorporated the recognition that networks are not derived from a random structure, and failures of nodes, whether from attacks or due to dependent correlations, are not uniformly random~\cite{albert2000error}. Hence, significant interest has developed in removal processes that model targeted attacks on the network using a centrality metric.  In the network science domain, the degree of network resilience is commonly measured based on the size of the giant component (i.e., the largest connected component in a given network), which gives a clear sense of how the network is connected with a portion of existing nodes even after a certain number of nodes or edges are removed.  Percolation theory has been used to model various processes on networks in the context of network failures or attacks, e.g., connectivity, routing, and epidemic spreading~\cite{Colbourn87, Najjar90}. 

\subsection{Software Diversity-based Cybersecurity} \label{subsec:software_diversity}

Many approaches have been explored to validate the usefulness of software or network diversity to ensure network security.
\citet{Chen16-safety} investigated the usefulness of software diversity to enhance security.  \citet{Huang14, Huang17} solved a software assignment problem by isolating nodes with the same software to minimize the effect of epidemic worm attacks. %This work considered not only the constraints of host functionality and software availability, but also the degree and effect of vulnerability to maximize system security based on the balance between these two key factors. 
\citet{Franz10} proposed an approach to introduce compiler-generated software diversity for a large scale network, aiming to create hurdles for attackers and eliminate any advantage of knowing the vulnerabilities of a single software. %An `App Store' with a diversification engine automatically generates a different version of every program, which is functionally identical whenever a download is requested. 
\citet{Homescu17} presented a large-scale automated software diversification to mitigate the vulnerabilities exposed by software monoculture. %This work gives the description of the implementation of the developed automated software diversity tool and proposed a generic method to measure the effectiveness of software diversity as a potential cure to remove code-reuse attacks. 
\citet{Yang08, Yang16} proposed a software diversity technique to combat sensor worms by solving a software assignment problem, given a limited number of software versions available. %Taking the key philosophy of software diversity to prolong system survivability, 
The authors used percolation theory to model the design features of software diversity to defend against sensor worms. %The authors also extended this work to investigate the effect of sensors with multiple software versions, rather than a single version in terms of network robustness under sensor worm attacks~\cite{Yang16}. \citet{Salako14} developed probability models to consider the commonalities in developing multiple versions of software and proposed alternative ways to consider software diversity in the presence of dependencies between the different versions of software.

\citet{Zhang01} developed a resilient, heterogeneous networking-based system where a single solution was common to increase interoperability. Recently, {\em network diversity} is proposed as a security metric to measure network resilience against zero-day attacks~\cite{Zhang16}. %This work designed and evaluated a suite of network diversity metrics such as a biodiversity-inspired metric based on the number of different resources, which has positively impacted in enhancing security. 
Inspired by the network diversity metrics~\cite{Zhang16}, \citet{Li18} further developed the network model and diversity metric based on vulnerability similarity, configuration constraints and multi-label hosts. 
%In order to prevent malware propagation, they tried to solve a software assignment problem and achieve the optimal software distribution.
\citet{hosseini18} mathematically analyzed the malware propagation under a network with six different types of nodes in an epidemic model. % following an epidemic mode, including susceptible,  exposed, infected, recovered, vaccinated and quarantine. 
%They conducted a sensitivity analysis on the number of software configurations by randomly distributing software over all nodes in the network. 
They proved a positive correlation between network security and the degree of network diversity.  \citet{prieto2019} proposed an optimal software assignment algorithm with multiple software packages to enhance network resilience under attacks.
%Based on their theoretic models and related results in experiments on eight real-world network topologies, their work demonstrates high impact of software diversity on realizing network resilience.

Although the above works discussed
%~\cite{Chen16-safety, Dezso02, Franz10, Hole15, Hole15-antifragile, Homescu17, Huang14, Huang17, Li18, Yang08, Yang16, Zhang16, Zhang01} \tjm{[<- I don't like these long list of references because there are always mistakes. For example, \cite{hosseini18}\cite{prieto2019} are not in the list but were mentioned and \cite{Dezso02}\cite{Hole15}\cite{Hole15-antifragile} are listed but not mentioned in this section. Were the references supposed to be mentioned or were the wrong ones listed?]} 
the concept of software diversity to ensure system security, their aim is to solve a software assignment problem by shuffling different types of software packages among nodes without changing the network topology.  Unlike the software assignment approach, we aim to generate an optimal network topology that is resilient against epidemic attacks while maximizing network connectivity.  The proposed software diversity metric is designed for each node to make a decision on whether to add or remove an edge based on the vulnerabilities on the attack paths reachable to the node~\cite{Chen07, Keramati13}. %\tjm{[Is there a good citation for attack path?]}

\section{System Model} \label{sec:system_model}
This section discusses our system model in terms of the network model, the node model, the attack model, and the defense model.

\subsection{Network Model} \label{subsec:network_model}
In this work, we are concerned with a special distributed network environment where each node belongs to a set of regional coordinators.  Examples include a software-defined network (SDN) where each node can be instructed by an SDN controller it belongs to~\cite{Kreutz15}, an edge computing Internet-of-Things (IoT) system with some high computing edge nodes available to perform high computing tasks~\cite{Li18-edge}, and a hierarchical mobile ad hoc network with decentralized controllers in charge of governing the nodes under their control~\cite{Cho08-decen-manet}.  Periodic information exchange between nodes and the regional coordinators is required to ensure seamless operations of the system. However, since each node's software diversity value, which is used to make a decision on edge adaptation (i.e., adding/removing edges), is computed locally by each node, a regional coordinator will only need to rank the software diversity values of neighbor nodes around a target node, and inform the target node of
which edges to add/remove based on the estimated ranks. 
Moreover, the ranking operation of the neighbor nodes around a target node is only periodically performed by a regional coordinator and will not require high communication overhead for each node to communicate with the regional coordinator. 

%\tjm{[Is $\mathbf{G}$ used elsewhere?]} , denoted by $\mathbf{G}$, is denoted by $\mathbf{V}$ where vertex $v_i$

A temporal network is an undirected network for which the topology evolution (or change) occurs due to node failures or nodes being compromised by attackers. In addition, the network may change its topology when adaptation strategies are performed by connecting between two nodes or disconnecting all the edges associated with compromised nodes to mitigate the spread of infection over the network. We denote the set of nodes in the network by nodes $i$'s, characterized by a set of attributes as shown in Section~\ref{subsec:node_model}.  %The set of edges between nodes is represented by $\mathbf{E}$ where 
An edge between nodes can be on and off depending on the dynamics caused by node failures, node recovery, or edge adaptations (i.e., an edge can be added or removed).  
We maintain an adjacency matrix $\mathbf{A}$ in order to keep track of direct or indirect connectivities (i.e., edges) between nodes where $a_{ij}=1$ indicates there exists an edge between nodes $i$ and $j$ while $a_{ij}=0$ indicates that no edge exists.

In order for each node to efficiently estimate its software diversity by considering the vulnerabilities of attack paths reachable to the node, it only considers neighboring nodes within $k$-hop distance from itself. We call it a node's {\em $k$-hop local network}. This local network is used for each node to estimate its software diversity value by considering the vulnerabilities of attack paths available within its local network.  

Although we will maintain the value of $k$ as a sufficiently small number (e.g., 1 or 2), it does not underestimate the vulnerabilities of possible attack paths because using smaller $k$ means taking a conservative perspective that an attacker is quite close enough such as being within the local network. For example, if an attacker wants to compromise a particular target node, it may try multiple attack paths where each attack path has a set of intermediate nodes. When the attack path is long, it means the vulnerability of the target node is low as the attacker needs to compromise all the intermediate nodes in order to finally compromise the target node.  However, when the path length is small, it does not necessarily decrease the attack vulnerability because the attacker is close to the target node.  
%The following is commented out because it is confusing for the reader to understand the detail of how a node can avoid being compromised or estimate its node vulnerability and also it does not convey important information
%In addition, for each node not to be compromised, when multiple attack paths exist, each node should not be successfully compromised by any of the compromised nodes. However, if considering a single attack path can fairly well estimate at least the `ranks' of non-vulnerabilities of nodes from possible attack paths, we don't need to consider all possible attack paths of each node, a process that would unnecessarily consume computational resource. 

We assume that software packages installed in each node and the associated vulnerabilities information are given to the regional coordinator in the initial network deployment period.  In addition, we assume that each node is also well informed about the software vulnerability information associated with the software packages installed in the neighboring nodes in its $k$-hop local network.  We assume that the changes of network topology are mainly made by node failures or network adaptations in this work.

Adding or removing an edge between two nodes requires secure communications between them. Even if they are within wireless range of each other but don't share a secret key for secure communications, they are not logically connected.  In this work, generating an optimal network topology which is resilient against epidemic attacks with maximum network connectivity is based on a logical network topology. 

\subsection{Node Model} \label{subsec:node_model}
Each node $i$ is characterized by its attributes as:
\begin{itemize}
\item A node $i$'s status on whether it is active or not, denoted by $na_i$, indicating whether it is alive ($=1$) or failed ($=0$), respectively; 
\item A node $i$'s status on whether the node is compromised ($=1$) or non-compromised ($=0$), denoted by $nc_i$; 
\item A node $i$'s software package installed, representing the diversified package or version of the same software providing the same functionality.  In this work, we adopt the well-known software diversification approach called {\em N-version programming}~\cite{Avizienis77, Avizienis85}.  This concept means that a software has multiple independent implementations while the different implementations of the software still can provide same functionalities while the implementations are different and naturally have different bugs or vulnerabilities. Following this concept, we model the node's software package installed denoted by $s_i$ with a limited number of software packages available, $N_s$, where $s_i$ is an integer, ranged in $[1, N_s]$;  
\item A node $i$'s degree of software diversity, $sd_i$, whose physical meaning is how different node $i$'s software package is from its neighbors.  The detail on the computation of node $i$'s software diversity are elaborated in Eq.~\eqref{eq:metric_sd}; and
\item A node $i$ has software vulnerability derived from the software package $s_i$ it is installed with, which is denoted by $sv_i$.
\end{itemize}
Based on the above five attributes, node $i$ is characterized by:
\begin{equation} \label{eq:node_attributes}
\mathbf{node} (i) = [na_i, nc_i, s_i, sd_i, sv_i]. 
\end{equation} 
If attacker $j$ targets vulnerable node $i$ (i.e., a node that has not been compromised before), which is one of its direct neighbors, the probability that node $j$ infects node $i$, denoted by $\beta_{ji}$, is estimated based on the probability that node $j$ can exploit the vulnerability of node $i$'s software package, $s_i$. We estimate this probability based on node $i$'s vulnerability to node $j$, estimated by~\cite{Hole15}:
\begin{equation} \label{eq:exploit_rho}
\beta_{ji} = 
\begin{cases} 
1  & \quad  \text{if } \sigma_j (\comg{s_i}) > 0 \text{;} \\
sv_{i} & \quad \text{otherwise,}
\end{cases}
\end{equation} 
where $\sigma_j$ is a vector of software packages attacker $j$ has learned about their security vulnerabilities.  For example, attacker $j$ knows the vulnerabilities of software package 1 and 3 among 5 packages available. It is denoted by $\sigma_j=[1, 0, 1, 0, 0]$. In this case, the sum of $\sigma_j$ indicates the number of software packages attacker $j$ knows the vulnerabilities of and so can exploit. Note that it is a dynamic value learned after node $j$ compromises node $i$ via reconnaissance even if their installed software packages are different, i.e., $s_i \neq s_j$. Here $sv_{i}$ refers to the vulnerability of software package $s_i$, which can be estimated based on the the degree of a Common Vulnerabilities and Exposures (CVE) with a Common Vulnerability Scoring System (CVSS) severity score~\cite{CVSS2018, CVE2018}. A node's mean vulnerability is simply obtained by the scaled mean vulnerability across multiple vulnerabilities in $[0, 10]$ where the maximum vulnerability score is 10 in CVSS. We normalize the value ranged in $[0, 1]$ as a real number.

\subsection{Attack Model} \label{subsec:attack_model}
This work deals with two stages of attack behaviors: An outside attacker before the node is compromised and an inside attacker after the node is compromised but undetected.

\vspace{1mm}
\noindent {\bf (1) Node Compromise by Epidemic Attacks}: We consider the so called {\em epidemic attack} which describes an attacker's infection behavior based on an epidemic model, called the SIR (Susceptible-Infected-Removed) model~\cite{Newman10}. That is, an outside attacker can compromise the nodes directly connected to itself, its direct neighbors, without access rights to their settings or files. Typical example scenarios include the spread of malwares or viruses. Botnets can spread malwares or viruses via mobile devices. A mobile device can misuse a mobile malware, such as a Trojan horse, thus acting as a botclient to receive commands and controls from a remote server~\cite{Mavoungou16}.  Further, worm-like attacks are popular in wireless sensor networks where the sensor worm attacker sends a message to exploit the software vulnerability in order to cause a crash or take control of sensor nodes~\cite{Yang08, Yang16}.  Attacker $j$ can compromise its direct neighbor $i$ when 
%the following two conditions are met:
%\begin{itemize}
%\item node $i$ has not been compromised by other attackers (i.e., $nc_i=0$); and 
%\item Node 
node $i$ uses a software package that attacker $j$ can exploit because the attacker knows the vulnerability of the software package. This case happens when $s_i$ is the same as $s_j$ or attacker $j$ learned $s_i$'s vulnerability in the past (i.e., $\mathbf{\sigma}_j(s_i)> 0$). When attacker $j$ is installed with a particular software package, $s_j$, we assume that attacker $j$ knows the vulnerability of its own software package, $s_j$. Attacker $j$ can learn the vulnerabilities of other software packages although it needs to commit more time and resources to obtain the information of their security vulnerabilities. Node $i$'s vulnerability by attacker $j$ based on these two cases is reflected in Eq.~\eqref{eq:exploit_rho}.
%\end{itemize}
When node $i$ is compromised, node $i$'s status is changed from `susceptible' to `infected' indicating that node $i$ is now an attacker. Then, node $i$ can infect other nodes and learn their software vulnerabilities, which are unknown to it. The attack procedures are described in Algorithm~8 of the appendix file.  

\vspace{1mm}
\noindent {\bf (2) Malicious Behavior of Compromised Nodes Undetected by the IDS}: Even if an intrusion detection system (IDS) is assumed to be placed in this work (see Section~\ref{subsec:defense_model} below), an attacker may not be detected by the IDS and the inside attacker can perform malicious behaviors such as packet dropping attacks (e.g., gray or black hole attacks), data exfiltration attacks, or denial-of-service (DoS) attacks to compromise the security goals in terms of loss of 
%security goals including 
confidentiality, integrity, and availability~\cite{Do15, Wood02}.

\subsection{Defense Model} \label{subsec:defense_model}
We assume that a system is equipped with an IDS, which detects infected (i.e., compromised) nodes.  When infected node $i$ is detected by the IDS, we model the detection probability with $\gamma$ representing the removal probability in the SIR model.  The response to the detected node will be performed by disconnecting all the edges connected to the detected attacker, which corresponds to removing the node from the system based on the concept of {\em site percolation}.  Note that the development of an IDS is beyond the scope of this work. We simply consider the IDS characterized by a false positive probability and a false negative probability, both of which have the value of $1-\gamma$. 

\section{Software Diversity based Adaptation Algorithm Design} \label{sec:software_diversity_strategies}
In this section, we describe our proposed software diversity based adaptation (SDA) algorithm design in detail. SDA uses software diversity as a key determinant to select edges to percolate (i.e., add or remove) for mitigating the spreading of compromised nodes by epidemic attackers and also to maximize the network connectivity for network resilience. 

\subsection{Software Diversity Metric} \label{subsec:diversity_vulnerability}

A node's vulnerability is commonly computed based on its software package installed~\cite{Hole15, Hole15-antifragile}. However, if the node is connected with many other nodes that are directly or indirectly connected, its potential vulnerability is not simply restricted by the vulnerability of its own software package. We use a broader concept of node vulnerability by embracing the vulnerabilities of attack paths reachable to each node.  To better capture the relationship between node vulnerability and network topology, we utilize an attack path $AP$ an attacker can take to successfully compromise a target node.  That is, in order to compromise the target node, the attacker needs to compromise all intermediate nodes on the attack path.  Hence, we estimate each node's software diversity value which refers to the probability that a node is robust against vulnerabilities from attack paths $AP$'s reachable to the node. 

\begin{figure*}[th!]
\centering
\subfigure[Software diversity estimation of node $i$]{
    \includegraphics[width =0.45\textwidth, height = 0.35\textwidth]{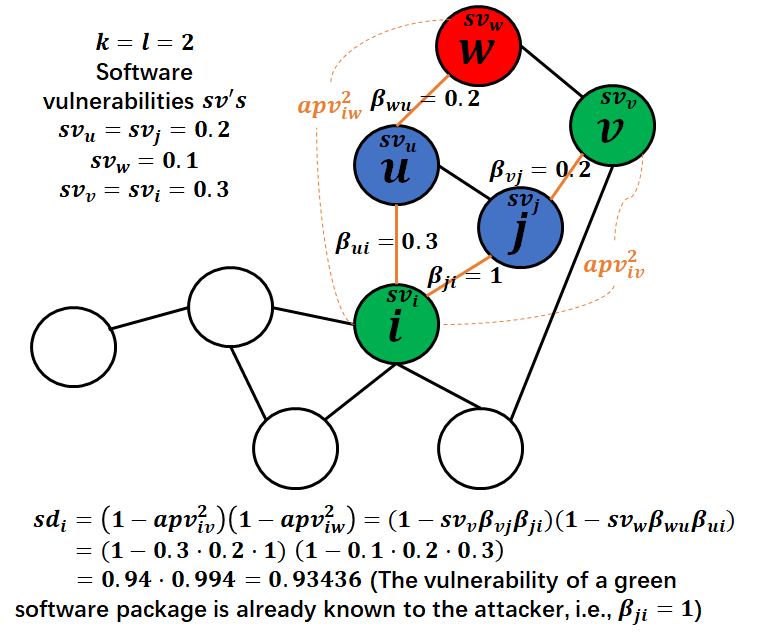}}
\subfigure[Edge adaptations based on software diversity differences]{
    \includegraphics[width=0.45\textwidth, height = 0.35\textwidth]{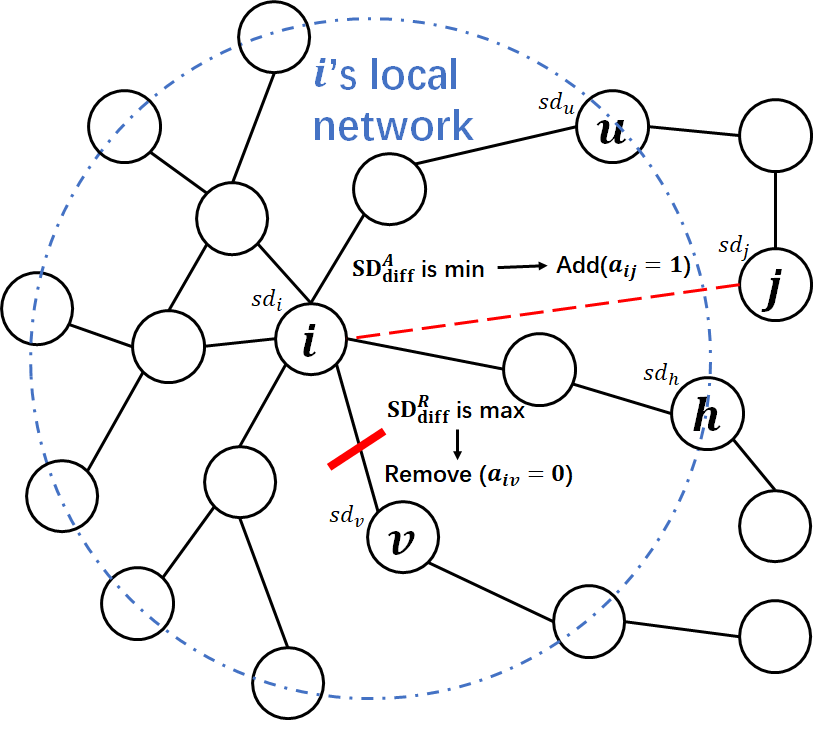}}
    \caption{Example of the software diversity-based adaptation strategies: (a) The estimation of node $i$'s software diversity value; and (b) The edge adaptation based on the software diversity difference in Eqs.~\eqref{eq:sd-diff-add} and~\eqref{eq:sd-diff-remove}.}
    \label{fig:sd-overview}
\end{figure*}

To this end, we consider the shortest paths (i.e., maximum $k$-hop distance paths) between boundary nodes (i.e., nodes in the boundary of a target node's local network) to a target node as attack paths.  In addition, to reduce the complexity of measuring each node's software diversity, we use a limited number of attack paths, denoted by $l$, where each path has at most $k$-hop distance.  Target node $i$'s software diversity, denoted by $sd_i (k, l)$, is obtained by:
\begin{equation}
\label{eq:metric_sd}
sd_i (k, l):=\prod_{j \in \mathbf{ap}_i}^l (1-apv^k_{ij}),
\end{equation}
where $\mathbf{ap}_i$ is a set of attack paths available to node $i$ ranked based on their highest vulnerability and $apv^k_{ij}$ is the vulnerability of an attack path from node $j$ to node $i$ with the maximum hop distance $k$.  In order to consider the maximum number of nodes associated with the attack paths, we consider disjointed attack paths from the boundary nodes (i.e., $j$'s) to node $i$.  %\jhc{Further, to consider the worst scenario with the highest vulnerability, we consider top $l$ most vulnerable attack paths.} 

%To mitigate potential bias in the estimation of software diversity metric, we maximize the total number of nodes considered in software diversity metric by only choosing attack paths ending at node and starting at entry points in the local network, where entry points are nodes that are not in any shortest paths from other nodes to the node. 

\subsection{Software Diversity based Bond Percolation for Network Adaptation}

\begin{algorithm}[!th]
\small{
\caption{Software Diversity-based Adaptation (SDA)}
\label{algo:SDA}
\begin{algorithmic}[1]
\State{$N \leftarrow$ The total number of nodes in a network}
\State{$\mathbf{DN} \leftarrow$ A vector containing the number of removed edges per node}
\State{$\mathbf{A} \leftarrow$ An adjacency matrix for a given network with element $a_{ij}$ for $i, j = 1, \ldots, N$}
\State{$\mathbf{S} \leftarrow$ A vector of software packages installed over nodes with element $s_i$ for $i = 1, \ldots, N$}
\State{$\mathbf{SV} \leftarrow$ A vector of the vulnerabilities associated with software packages}
\State{$\mathbf{SD} \leftarrow$ A vector of software diversity values, $sd_i$ for all nodes $i = 1, \ldots, N$}
\State{$\mathbf{PV} \leftarrow$ A vector of maximum path vulnerabilities for all nodes $i$ where $pv_i$ refers to the maximum attack path vulnerability where the path consists of at most $k-1$-hop distance from node $i$} \State{$k \leftarrow$ A hop distance given in a node's local network}
\State{$l \leftarrow$ A maximum number of attack paths considered for estimating a node's software diversity}
\State{$\rho \leftarrow$ A threshold referring to the fraction of edges to be removed when $\rho<0$ and added when $\rho>0$}
\State{$\mathbf{A}' \leftarrow$ An adjacency matrix after edges are adapted in Step 1.}
\State{$\mathbf{A}'' \leftarrow$ An adjacency matrix after edges are adapted in Step 2.}
\\
\State{$\mathbf{A}''= \mathbf{SDA} (N, \mathbf{DN}, \mathbf{A}, \mathbf{S}, \mathbf{SV}, k, l, \rho)$}
\\
\State{{\bf Step 1:} $\mathbf{A}' = \mathbf{SDBA} (N, \mathbf{DN}, \mathbf{A}, \mathbf{S}))$} \Comment{Remove edges between two nodes with the same software package based on Algorithm~1 of the appendix file).}  
\\
\State{{\bf Step 2:} Add/remove edges locally based on the ranks of the software diversity differences estimated in Eqs.~\eqref{eq:sd-diff-add} and~\eqref{eq:sd-diff-remove}} (Algorithms~4 and 5 of the appendix file)
\State{$\mathbf{A^*}\leftarrow \mathbf{(A'+I)}^{2k}$ where $a^*_{ij}$ is 1 when nodes $i$ and $j$ belong to each other's local network or their neighbors' local networks; 0 otherwise.}
\State{$\mathbf{SD} \leftarrow$ A vector of software diversity where each element,  $sd_i (k, l)$, refers to node $i$'s software diversity value when at most $l$ number of attack paths are considered where each attack path has at most $k$-hop length.}
\State{$\mathbf{PV} \leftarrow$ A vector of estimated attack path vulnerabilities associated with each node.}
%\State{$\mathbf{count}\leftarrow$ A counter for the number of adapted edges per node, initialized at 0.}
\State{$\mathbf{candidate}\leftarrow$ A set of edge candidates} \Comment{Algorithms~4 and 5 of the appendix file.}
%\State{$\mathbf{MAX}\leftarrow$ estimated local maximum SD differences for each node in the network, initialized as 0. \jhc{not used}}
%\State{$\mathbf{MIN}\leftarrow$ estimated local minimum SD differences for each node in the network, initialized as 1. \jhc{not used}}
\State{$\mathbf{T}^{local}, T^{global} = \mathbf{setEAB} (\mathbf{DN},\mathbf{A}',\rho)$}\Comment{Set edge adaptations budget based on Algorithm~3 of the appendix file.}
\If{$\rho>0$}
\State{$\mathbf{candidate} = \mathbf{GEAC}(\mathbf{A}',\mathbf{SD},\mathbf{SV},\mathbf{S},\mathbf{PV},\mathbf{T}^{local}$)} \\
\Comment{Algorithm~4 in the appendix file.}
\Else
\State{$\mathbf{candidate}$ = $\mathbf{GERC}(\mathbf{A}',\mathbf{SD},\mathbf{SV},\mathbf{S},\mathbf{PV},\mathbf{T}^{local}$)} \\
\Comment{Algorithm~5 in the appendix file.}
\EndIf
\State{$\mathbf{A}''=\mathbf{AdaptNT}(\mathbf{A}',\mathbf{candidate}, \mathbf{T}^{local},T^{global},\rho$)} \\ \Comment{Algorithm~6 in the appendix file.}

\State{$\mathbf{return}\  \mathbf{A}''$}  
\end{algorithmic}	
}
\end{algorithm}
The design objective of SDA is to decide which edges to add or remove in order to maximize the size of the giant component (i.e., the largest network cluster in a network) for maintaining network connectivity and to minimize the fraction of nodes being compromised due to epidemic attacks with minimum defense cost defined in Section \ref{subsec:metrics}.

We have two tasks to determine which edges to remove or add as follows:
\begin{enumerate}
\item Estimate the gain or loss as a result of removing or adding an edge based on the difference between a node's current software diversity value and its expected software diversity value after the edge adaptation made between nodes $i$ and $j$. To determine if adding an edge between nodes $i$ and $j$ is beneficial, we compute the software diversity difference by:
\begin{equation}
SD_{\mathrm{diff}}^{A} (i, j) = (sd_i - sd'_i) + (sd_j - sd'_j), 
\label{eq:sd-diff-add}
\end{equation}
where $sd_i = sd_i (k, l)$ and $sd_j = sd_j (k, l)$ for simplicity and $sd_i (k, l)$ and $sd_j (k, l)$ are defined in Eq.~\eqref{eq:metric_sd}.  $sd'_i$ and $sd'_j$ are the expected software diversity values of nodes $i$ and $j$ after an edge is added. The most promising candidate edge to be added should be an edge with the lowest $SD_{\mathrm{diff}}^A (i, j)$.  $sd'_i$ is simply obtained by
\begin{equation}
sd'_i = sd_i (1-sv_{i} pv_j),   
\end{equation}
where $sv_{i}$ is the software vulnerability of the software package installed in node $i$ (i.e., $s_i$) and $pv_j$ is the attack path vulnerability from node $j$ to the boundary node in node $j$'s local network.  That is, $sv_{i} pv_j$ is the same as $apv_{ij}$ in Eq.~\eqref{eq:metric_sd} (where we omitted $k$ for simplicity). $sd'_j$ can also be similarly obtained.
%For an edge to be removed, we similarly switched the order to obtain the difference because removing an edge is more likely to increase a node's software diversity due to reduced vulnerability. 
To determine if removing the edge between nodes $i$ and $j$ is beneficial, we compute the software diversity difference by:
\begin{equation}
SD_{\mathrm{diff}}^{R} (i, j) = (sd'_i - sd_i) + (sd'_j - sd_j), 
\label{eq:sd-diff-remove}
\end{equation}
where $sd'_i$ is computed by:
\begin{equation}
sd'_i = sd_i/(1-sv_{i} pv_j).
\end{equation}
Here the division by $(1-sv_{i} pv_j)$ represents the extent of reducing the vulnerability by removing an edge between nodes $i$ and $j$ based on Eq.~\eqref{eq:metric_sd}.  $sd'_j$ is similarly obtained like $sd'_i$ above. The most promising candidate edge to be removed should be an edge with the highest $SD_{\mathrm{diff}}^R (i, j)$.  See Algorithm~4 (Generates Edge Addition Candidates or GEAC) and Algorithm~5 (Generates Edge Removal Candidates or GERC) of the appendix file that provides the detail on generating edge candidates for edge addition and removal, respectively.
\item Estimate how many edges each node can adapt, i.e., remove or add. Based on the rationale that high centrality nodes (e.g., high degree) may expose high vulnerability in terms of security and network connectivity, we minimize the difference between the maximum degree and minimum degree by adding more edges to nodes with lower degree while deleting edges to the nodes with higher degree. Based on this principle, we develop a heuristic method to estimate how many edges should be adapted per node.  See Algorithm~3 (Set Edge Adaptations Budget or SetEAB) of the appendix file for detail.
\end{enumerate}
Algorithm~\ref{algo:SDA} describes our proposed software diversity-based adaption (SDA) algorithm in detail.  It executes Algorithm~1 of the appendix file in Step 1 on line 16 to remove edges between two nodes with the same software package. It makes the decision to add/remove edges locally based on the ranks of the software diversity differences estimated in Step 2 based on Eqs.~\eqref{eq:sd-diff-add} and~\eqref{eq:sd-diff-remove}, with the objective to best satisfy both security (i.e., minimum or no impact by epidemic attacks) and performance (i.e., a sufficient level of network connectivity) requirements. 
%There exists an optimal number of edges that can best satisfy both security (i.e., minimum or no impact by epidemic attacks) and performance (i.e., a sufficient level of network connectivity).  
%In this work, we investigate this optimal network density by varying the fraction of edges to be adapted, $\rho$, ranged in $[-1, 1]$ by either removing 100\% to 0\% with respect to the total number of edges remaining after Step 1 -- SDBA in  Algorithm~\ref{algo:SDA-step-1} or adding 0\% to 100\% from the total number of edges lost in Step 1. 

Fig.~\ref{fig:sd-overview} illustrates the SDA algorithm execution with an example network where distinct software packages are marked with distinct colors. Fig.~\ref{fig:sd-overview} (a) illustrates how node $i$ estimates its software diversity value when $k=l=2$. Fig.~\ref{fig:sd-overview} (b) illustrates how node $i$ determines whether to add or remove edges based on the software diversity differences, $SD_{\mathrm{diff}}^{A}$ and $SD_{\mathrm{diff}}^{R}$, based on Eqs.~\eqref{eq:sd-diff-add} and~\eqref{eq:sd-diff-remove}, respectively.

\section{Experimental Setup} \label{sec:exp-setup}
In this section, we describe performance metrics, counterpart baseline schemes against which our proposed SDA algorithm (i.e., Algorithm~\ref{algo:SDA}) is compared for performance comparison, and simulation environment setup for performance evaluation.  %Finally, we demonstrate the experimental results and discuss the overall trends observed from the results. 

\subsection{Performance Metrics} \label{subsec:metrics}
We use the following performance metrics:
\begin{itemize}
\item {\bf Software diversity ($SD$)}: This metric measures the mean software diversity for all nodes in a network. Since node $i$'s software diversity, i.e., $sd_i$, is computed based on Eq.~\eqref{eq:metric_sd}, the mean software diversity for all nodes in the network is obtained by:
\begin{equation} \label{eq:metric_average_sd}
SD = \frac{\sum_{i=1}^N sd_i}{N}.
\end{equation}
Recall that $k$ is used to determine node $i$'s local network and thus is the maximum possible hop distance from node $i$ to all other neighboring nodes in its local network. Higher software diversity is more desirable to ensure high system security.  
\item {\bf Size of the giant component ($S_g$)}: This metric captures the degree of network connectivity composed of non-compromised (uninfected), active nodes in a network. $S_g$ is computed by:
\begin{equation} \label{eq:metric_giant_component}
S_g = \frac{N_g}{N},
\end{equation}
where $N$ is the total number of nodes in the network and $N_g$ is the number of nodes in the giant component. Higher $S_g$ is more desirable, implying higher network resilience in the presence of epidemic attacks.
\item {\bf Fraction of compromised nodes ($P_c$)}: This metric measures the fraction of the number of compromised nodes due to epidemic attacks over the total number of nodes in a network.  This includes both currently infected (not detected by the IDS) and removed (previously infected and detected by the IDS) nodes.  $P_c$ is computed by:
\begin{equation} \label{eq:metric_vulnerable_component}
P_c = \frac{N_c}{N},
\end{equation}
where $N_{c}$ represents the total number of compromised nodes after epidemic attacks on a network (i.e., the original network under No-Adaptation and an adapted network under all adaptation schemes). See Section \ref{subsec:comparing_schemes} for a listing of counterpart baseline schemes against which our proposed SDA algorithm is compared for a comparative performance analysis. 
\item {\bf Defense cost ($D_c$)}: This metric measures the defense cost associated with the following defense strategies employed by an adaptation scheme: (1) edge adaptations (i.e., adding or removing edges) to isolate detected attackers (or compromised nodes) by the IDS; (2) edge adaptations to maximize software diversity by each node based on the value of the software diversity metric in Eq.~\eqref{eq:metric_sd}; and (3) shuffling operations based on the fraction of nodes whose software package is randomly shuffled over the total number of nodes. $D_c$ is computed by:
\begin{eqnarray} \label{eq:metric_dc}
D_c & = & \frac{\mathrm{sum}(|\mathbf{A}-\mathbf{B}|)}{\mathrm{sum}(\mathbf{A}+\mathbf{B})} + \frac{N_{SF}}{N}
\end{eqnarray}
%\jhc{change the denominator to the number of edges in a fully connected network where all are connected to all; this would work when the shuffling happens multiple times for future work in MTD}
In the first term, the numerator refers to the differences of edges between the adjacency matrix of an original network $\mathbf{B}$ and that of an adjusted network $\mathbf{A}$ after edges adaptations are made. The denominator is the sum of the additive two matrices. In the second term, $N_{SF}$ is the number of nodes whose software packages are shuffled and $N$ is the total number of nodes. Note that when a node's software package is shuffled but stays with its original software package, it is excluded from counting toward $N_{SF}$. This shuffling cost is estimated only when shuffling a software package is used such as random graph coloring, which is compared against our proposed SDA scheme in our work.  Lower defense cost is more desirable. 
\end{itemize}

\subsection{Counterpart Baseline Schemes for Performance Comparison} \label{subsec:comparing_schemes}
In this work, we compare the performance of our proposed SDA scheme against No-adaptation (No-A), Random adaptation (Random-A), and Random graph coloring (Random-Graph-C) counterpart baseline schemes for a comparative performance analysis. 

Our SDA scheme uses the software diversity-based metric in Eq.~\eqref{eq:metric_sd} to select an edge to remove or add based on the concept of bond percolation, as discussed in Section \ref{subsec:percolation_theory}. To be specific, SDA first removes all edges between two connected nodes with the same software package as shown in Step 1 of Algorithm~\ref{algo:SDA} (i.e., executing Algorithm~1 in the appendix file).  Then SDA decides a set of edges to be added or removed given $\rho$ (the percentage of edges to be added if $\rho>0$ or to be removed if $\rho<0$) as shown in Step 2 of Algorithm~\ref{algo:SDA}. The effect of $\rho$ on performance will be analyzed in Section~\ref{subsec:results_sensitivity} to identify the optimal $\rho$ value that can best balance security and network connectivity.
We experiment with various $\rho$ values in the range of $[-1, 1]$ where $-1$ means removing all edges (such that no edges exist in the network) and 1 means fully restoring edges removed from Step 1. For example, SDA with $\rho = 1$ means fully restoring edges lost from Step 1 while SDA with $\rho = 0$ refers to only executing Step 1 (removing edges between two nodes with the same software package).  SDA with $\rho = 0.6$ means only restoring 60\% of the edges lost in Step 1 while SDA with $\rho = -0.6$ means removing 60\% of edges in the network after Step 1.  What edges to remove or add (see Step 2 of Algorithm~\ref{algo:SDA}) significantly affects network security and resilience.
% the following is commented out because this is out of place. We can being it back in Section 6 if necessary
%To provide a lightweight solution, we also use limited knowledge for each node to estimate its software diversity based on the vulnerabilities of the attack paths reaching the node. We examined the performance of the variants of the SDA by varying the number of attack paths considered (i.e., $l$ number of attack paths) and the length of each attack path (i.e., maximum $k$-hop distance length in each attack path) in the proposed SDA.  
%We use the notation of SDA-$k$-$l$-$\rho$, which is the SDA scheme when the maximum hop distance in each attack path is $k$, the number of attack paths considered is $l$, and the fraction of edges restored after Step 1 is $\rho$. 

Below we briefly discuss the three counterpart baseline schemes to be compared against our proposed SDA schemes: 
\begin{itemize}
\item {\bf No-adaptation (No-A)}: This represents the case in which no adaptation is applied, thus showing the effect of attacks on the performance of the original network. However, we allow an IDS to detect attackers.
% the following is commented out because saying seeding attackers is some experiment detail which should not be mentioned when this baseline scheme is introduced
%after a set of seeding attackers is spread out to compromise other nodes that are directly connected to them in the network. 
When the IDS detects compromised nodes with probability $\gamma$, all edges connected to the detected attacker will be disconnected in order to isolate the attackers, ultimately resulting in mitigating the spread of compromised nodes in the network. Therefore, when No-A is used, the adaptation cost can be high because the number of edges disconnected is affected by the network topology, which is one of the key factors impacting the degree of network vulnerability.
\item {\bf Random adaptation (Random-A)}: This scheme first removes an edge between two nodes with the same software package (i.e., executing Algorithm~1 in the appendix file) and then randomly adds edges between nodes with a different software package (see Algorithm 7 in the appendix file).  In this scheme, we add the same number of edges lost due to the execution of Step 1. 
\item {\bf Random graph coloring  (Random-Graph-C)}: This scheme uses a simple rule for each node to shuffle its software package with the least common software package without changing any network topology. As a special case, when a node has many neighbors, it may choose the least common software package of those used among its neighbors. It may occur that a node shuffles to its original software package. In such a case, when the shuffled software package is the same as the original software package, we do not count it toward the shuffling cost in Eq. \eqref{eq:metric_dc}. We treat this scheme as an adaptation scheme because it also involves changing a configuration of its software by using a different implementation although it does not make any change to the network topology.
\end{itemize}

The pseudocode for SDA is presented in Algorithm~\ref{algo:SDA} and that for Random-A is described in Algorithm 7 of the appendix file.  In our experiment, we compare the performance of No-A, Random-A, Random-Graph-C, and three variants of SDA with three different thresholds $\rho$ in terms of the 4 performance metrics discussed in Section~\ref{subsec:metrics}. We treat Random-A, Random-Graph-C, and the SDA schemes as adaptation schemes while No-A is treated as a baseline scheme without adaptation.

\begin{table}[t!]
\caption{Key design parameters, their meanings, and their default values.}
\vspace{-2mm}
\begin{tabular}{|P{1cm}|p{5.5cm}|P{0.9cm}|}
\hline
\multicolumn{1}{|c}{Param.} & \multicolumn{1}{|c}{Meaning} & \multicolumn{1}{|c|}{Value} \\
\hline
$N$ & Total number of nodes in a network & 1000 \\
$p$ & Connection probability between pairs of nodes in a ER network & 0.025 \\
$\gamma$ & Intrusion detection probability & 0.95 \\
$k$ & The upper bound of hops considered in calculating software diversity $SD_{k,l}^{i}$ & [1,2] \\
$l$ & The upper bound of $\#$ of paths considered in calculating software diversity $SD_{k,l}^{i}$ & 1 \\
$n_r$ & Number of simulation runs & 100 \\
$N_s$ & Number of software packages available & [3,7] \\
$P_a$ & percentage of attackers in a network & [10,30\%]\\
$\rho$ & Threshold of fraction of edges adapted & $[-1,1]$  \\
$\mathbf{SV}$ & \vspace{-10mm}A vector of vulnerabilities associated with software packages which are selected based on the uniform distribution with the range in $(0, 0.5]$ (i.e., $U (0, 0.5]$).  For the maximum 7 different software packages, the $\mathbf{SV}$ of the corresponding vulnerabilities are used. & \scalebox{0.75}{$ \left[\begin{array}{c} 0.41 \\ 0.35 \\ 0.48 \\ 0.22\\ 0.16 \\ 0.19 \\ 0.12 \end{array}\right]^T$} \\
\hline
\end{tabular} 
\label{table:param_default_values}
\end{table}
%[0.41,0.35,0.48,0.22,0.16,0.19,0.12]

\subsection{Environment Setup}

%fig8
\begin{figure*}[!ht]
  \centering
    %\subfigure{
    %\includegraphics[width=0.12\textwidth, height=0.039\textwidth]{figs/fig8/legend.png}}\hspace{50em}\vspace{-0.5em}
    %\setcounter{subfigure}{0}
  \subfigure[Dense Network]{
    \includegraphics[width=0.251\textwidth, height=0.19\textwidth]{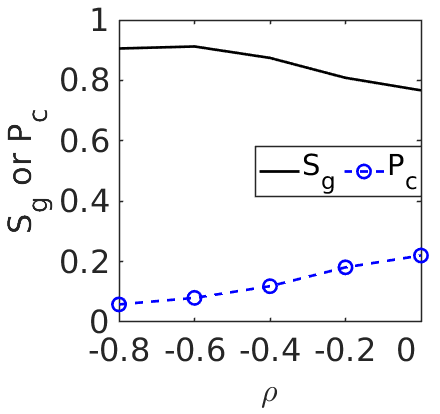}}
  \subfigure[Medium Network]{
    \includegraphics[width=0.251\textwidth, height=0.19\textwidth]{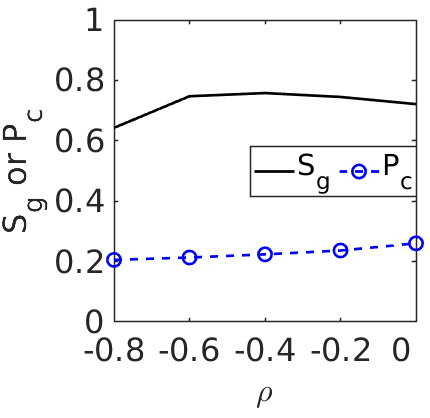}}
  \subfigure[Sparse Network]{
    \includegraphics[width=0.251\textwidth, height=0.19\textwidth]{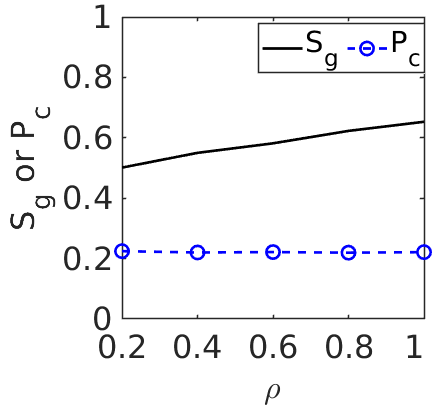}}
    \caption{Effect of $\rho$ (fraction of edges to be adapted) on performance of SDA in terms of the size of a giant component ($S_g$) and the fraction of compromised nodes ($P_c$). The optimal $\rho$ for the SDA scheme with respect to $S_g$ and $P_c$ in dense, medium dense, and sparse networks are identified as $\rho = -0.6, \rho = -0.4$ and $\rho = 1$, respectively.}
\label{fig_sensitivity}
\end{figure*}

\subsubsection{Parameters and Data Collection}

Table~\ref{table:param_default_values} summarizes the key parameters, their meanings, and their default values used in this work. We use the average of the performance measures collected based on 100 simulation runs.
In the experiment, we examine the effect of the following key design parameters on performance: (1) attack density (i.e., percentage of attackers); and (2) the number of software packages available.  For the ER network, we also study the effect of the network connection probability on performance in Section C.3 of the appendix file.  

\subsubsection{Network Topology Datasets}

We setup 4 different undirected networks to evaluate the proposed work: (1) a sparse network from an observation of the Internet at the autonomous systems level~\cite{snapnets}; 
(2) a medium dense network derived from an Enron email network~\cite{snapnets};  
(3) a dense Facebook ego network~\cite{snapnets}; and (4) an Erd{\"o}s-R{\'e}nyi (ER) random network~\cite{Newman10}. 
The network topologies and their degree distributions are shown in Figs.~1 and 2 of the appendix file.  Except for the medium dense network, we use the original network topologies. For the medium dense network, in order to derive a network of comparable size with the other networks (the Enron email network has 36,692 nodes and 183,831 edges) we generate the medium dense network with 985 nodes and 7,994 edges by taking the following procedures: (i) Rank all nodes in the Enron email network by degree in a descending order; (ii)  identify the medium dense network as the largest connected component of the induced subgraph, consisting of nodes with ranks from 501 to 1500, from the original graph.

%\tjm{[I'm wondering if it makes sense to split this section here? So Section 6 after this.]}

\subsubsection{Optimal Parameter Settings Used for SDA}
\label{subsec:results_sensitivity}
{\bf Fraction of edges to be adapted ($\rho$)}: 
We have conducted a sensitivity analysis of $\rho$ for the SDA scheme in terms of maximizing the size of a giant component ($S_g$) for network resilience without overly increasing the fraction of compromised nodes ($P_c$) for network security. 
As shown in Fig.~\ref{fig_sensitivity}, the optimal $\rho$ for the SDA scheme with respect to $S_g$ and $P_c$ in dense, medium dense and sparse networks have been identified as $\rho = -0.6, \rho = -0.4$ and $\rho = 1$, respectively. Due to space constraints, we have conducted the sensitivity analysis of $\rho$ for the ER random network in Appendix C.2 of the appendix file, from which we have observed the optimal $\rho$ with respect to $S_g$ and $P_c$ for the ER random network is $-0.6$. In summary, the optimal values of $\rho$ are observed at $-0.6$, $-0.4$, $1$, and $-0.6$ for dense, medium dense, sparse, and ER random networks, respectively.  
%\subsection{Effect of Network Density on Security Vulnerability and Network Connectivity}
%\label{subsec:results_sensitivity}

%In this work, we analyze the effect of $\rho$ (fraction of edges to be adapted) on the performance of our proposed SDA scheme and identify the best setting of $\rho$ to maximize the size of the giant component ($S_g$) representing network connectivity and minimize the fraction of compromised nodes ($P_c$) representing security vulnerability. This critical tradeoff between security and network resilience (connectivity) is significantly affected by the network characteristics.  By considering the network density of each network tested in this study, we conducted the sensitivity analysis under varying $\rho$ with the following experimental setup: (1) For the dense network, we used the fraction of initial  attacks with $P_a=0.05$ and the number of software packages available, $N_s=5$; (2) For the medium dense network, we used $P_a=0.2$ and $N_s=5$; and (3) For the sparse network, we used $P_a=0.2$ and $N_s=5$.  The reason of using $P_a = 0.05$ under the dense network is because the dense network is already highly vulnerable to epidemic attacks due to its high network density. To demonstrate more clear and interesting trends, we used a lower range of $P_a$ for the sensitivity analysis varying $P_a$ under the dense network.
%\tjm{[Is this explained someone? Why different for dense than others?]} 

\vspace{1mm}
\noindent {\bf The number of maximum attack paths ($l$) and the maximum hop distance in each attack path ($k$)}:  
The network type (i.e., dense, medium dense, sparse, or ER random) affects node density which in turn can affect the optimal setting of $l$ and $k$ under which SDA can best achieve both security (i.e., a low fraction of compromised nodes) and network resilience (i.e., a large size of the giant component). 
We have conducted a sensitivity analysis of $l$ or $k$ on performance of the SDA scheme in all four types of networks. Due to space constraints, we put the sensitivity analysis of $l$ and $k$ on performance of SDA in Sections D and E of the appendix file.  
In summary, for dense, medium dense, and ER random networks, we have selected $k=1$ and $l=1$ to calculate software diversity $sd_i (k, l)$ for each node in the network because we have observed no significant performance improvement with $k> 1$ and $l>1$.  For the sparse network, we have not observed high sensitivity when $l> 1$. However, for $k$, we have observed that SDA performs the best when $k=2$ 
with $\rho=1$. Thus, we have selected $k=2$ and $l=1$ for the sparse network. 

\section{Experimental Results and Analysis} \label{sec:exp-result-analysis}
In this section, we present the experimental results for a comparative performance analysis of the proposed SDA scheme against the counterpart baseline schemes and provide physical interpretations of the results.
In our experiment, we compare 6 schemes: (1) Non-adaptation (No-A); (2) Random adaptation (Random-A); (3) Random graph coloring (Random-Graph-C); (4) SDA with $\rho=0$; (5) SDA with $\rho=1$; and (6) SDA with optimal $\rho$. See Section~\ref{subsec:comparing_schemes} for more detail on how each scheme is implemented. 
The 6th scheme "SDA with optimal $\rho$" is network-type dependent. As discussed earlier in Section \ref{subsec:results_sensitivity}, the optimal values of $\rho$ are observed at $-0.6$, $-0.4$, $1$, and $-0.6$ for dense, medium dense, sparse, and ER random networks, respectively.

Initially a set of attackers is randomly and uniformly distributed to the network based on the percentage of attackers parameter $P_a$ and all such attackers perform epidemic attacks as described in Section~\ref{subsec:attack_model}. See Algorithm 8 of the appendix file for detail on how the attackers perform epidemic attacks.  Below we only report the experimental results under dense, medium dense, and sparse networks. The experimental results under the ER random network are reported in Section C.3 of the appendix file due to space constraints.

\begin{figure*}[!ht]
  \centering
  \subfigure{
    \includegraphics[width=0.65\textwidth, height=0.02\textwidth]{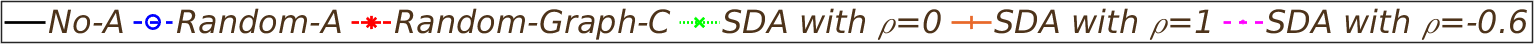}}\vspace{-1em}
   
   \setcounter{subfigure}{0}

  \subfigure[Fraction of compromised nodes ($P_c$)]{
    \includegraphics[width=0.25\textwidth, height=0.19\textwidth]{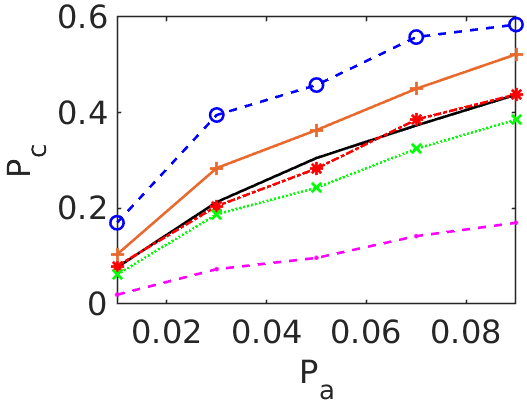}} %\hspace{-0.85em}
  \subfigure[Size of a giant component ($S_g$)]{
    \includegraphics[width=0.24\textwidth, height=0.19\textwidth]{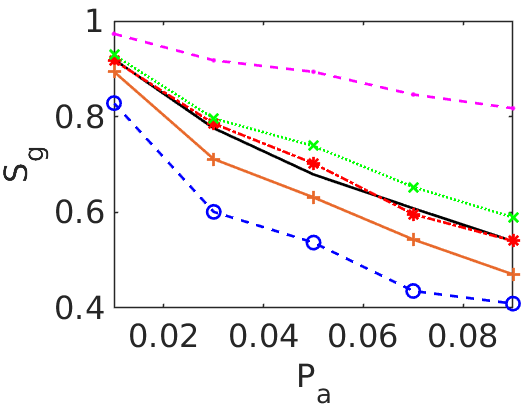}}\hspace{-0.85em}
  \subfigure[Software diversity ($SD$)]{
    \includegraphics[width=0.24\textwidth, height=0.19\textwidth]{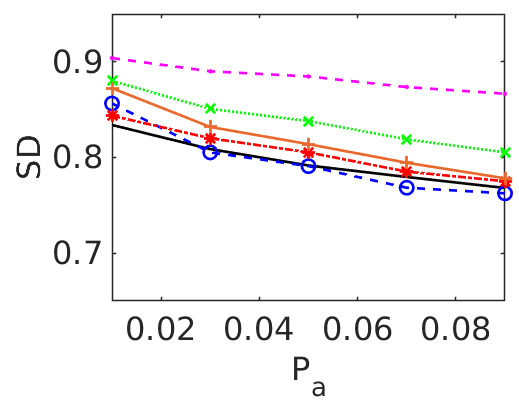}}
  \subfigure[Defense cost ($D_c$)]{
    \includegraphics[width=0.24\textwidth, height=0.19\textwidth]{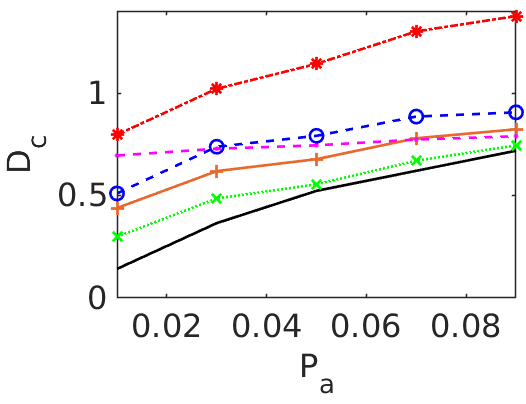}}\hspace{-0.85em}  
    \caption{Effect of varying the fraction of attackers ($P_a$) under a dense network.}
\label{fig_p_a_dn}
\end{figure*}
%fig 3
\begin{figure*}[!ht]
  \centering
  \subfigure{
    \includegraphics[width=0.65\textwidth, height=0.02\textwidth]{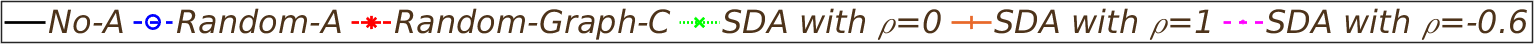}}\vspace{-1em}
   
   \setcounter{subfigure}{0}

  \subfigure[Fraction of compromised nodes ($P_c$)]{
    \includegraphics[width=0.25\textwidth, height=0.19\textwidth]{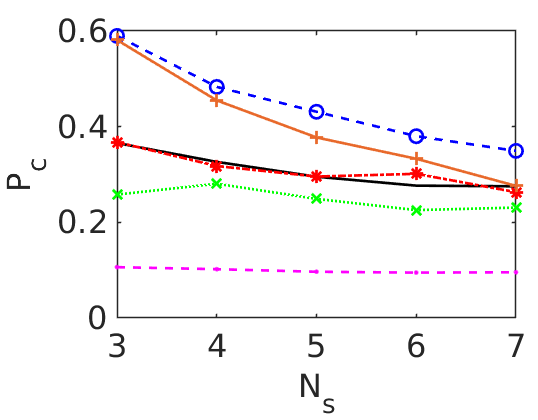}}\hspace{-0.5em}
  \subfigure[Size of a giant component ($S_g$)]{
    \includegraphics[width=0.24\textwidth, height=0.19\textwidth]{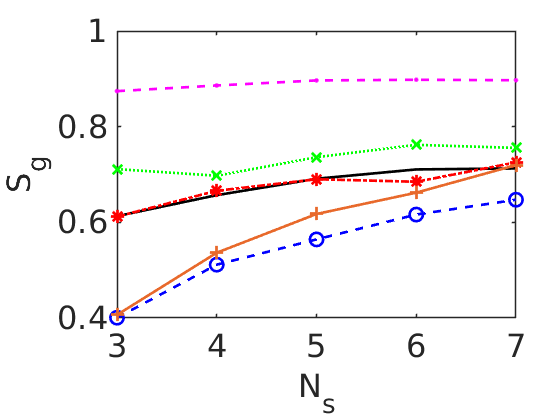}}\hspace{-0.85em}
  \subfigure[Software diversity ($SD$)]{
    \includegraphics[width=0.24\textwidth, height=0.19\textwidth]{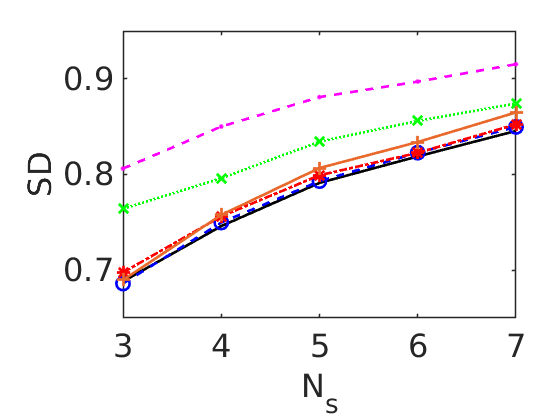}}
      \subfigure[Defense cost ($D_c$)]{
    \includegraphics[width=0.24\textwidth, height=0.19\textwidth]{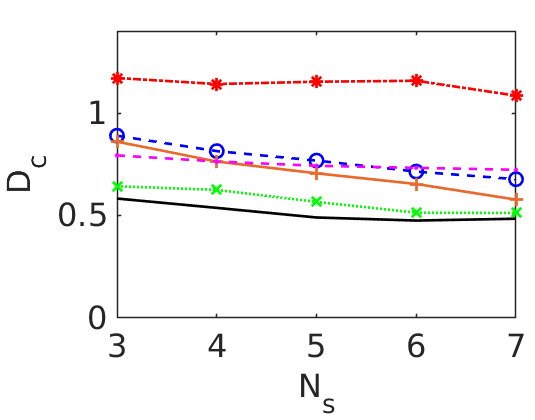}}\hspace{-0.85em}  
    \caption{Effect of the number of software packages ($N_s$) under a dense network.}
\label{fig_n_s_dn}
\end{figure*}

\subsection{Comparative Performance Analysis under a Dense Network}
%fig 4
\begin{figure*}[!ht]
  \centering
  \subfigure{
    \includegraphics[width=0.65\textwidth, height=0.02\textwidth]{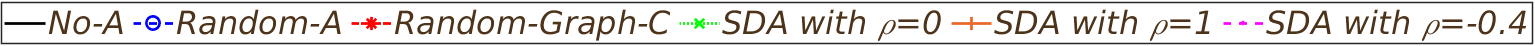}}\vspace{-1em}
   
   \setcounter{subfigure}{0}
  \subfigure[Fraction of compromised nodes ($P_c$)]{
    \includegraphics[width=0.25\textwidth, height=0.19\textwidth]{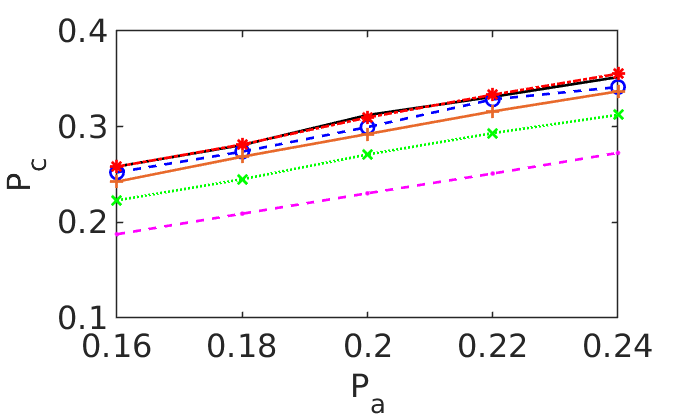}}\hspace{-0.5em}
  \subfigure[Size of a giant component ($S_g$)]{
    \includegraphics[width=0.24\textwidth, height=0.19\textwidth]{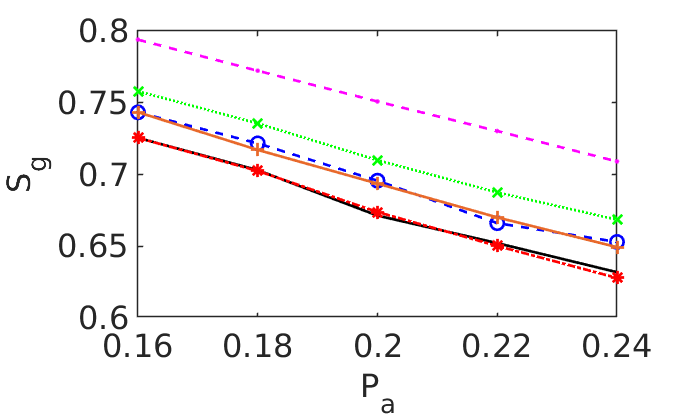}}\hspace{-0.85em}
  \subfigure[Software diversity ($SD$)]{
    \includegraphics[width=0.24\textwidth, height=0.19\textwidth]{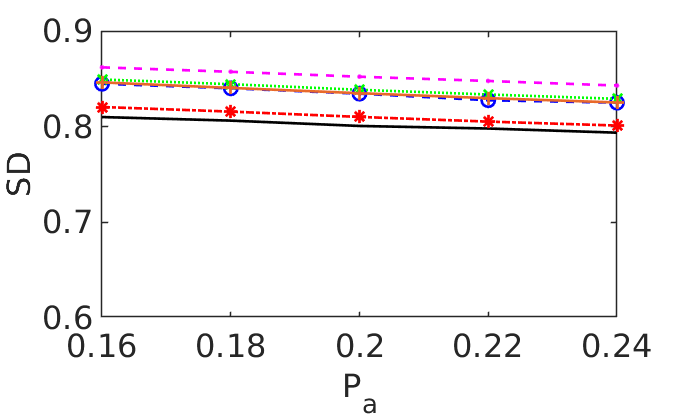}}\hspace{-0.85em}
      \subfigure[Defense cost ($D_c$)]{
    \includegraphics[width=0.24\textwidth, height=0.19\textwidth]{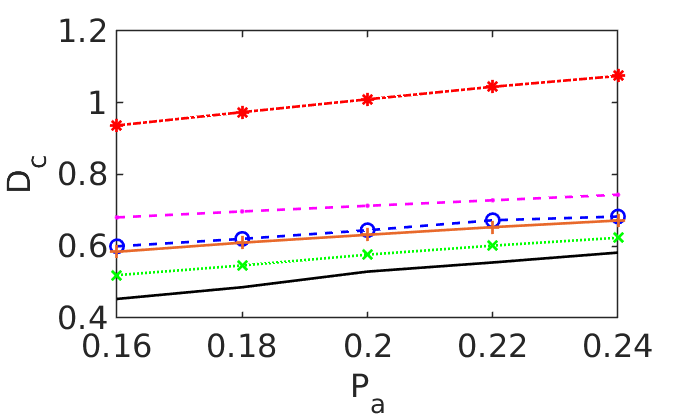}}\hspace{-0.85em} 
    \caption{Effect of varying the fraction of attackers ($P_a$) under a medium network.}
\label{fig_p_a_mn}
\end{figure*}

%fig 5
\begin{figure*}[!ht]
  \centering
  \subfigure{
    \includegraphics[width=0.65\textwidth, height=0.02\textwidth]{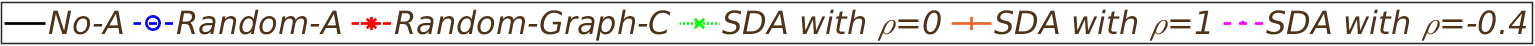}}\vspace{-1em}
   
   \setcounter{subfigure}{0}
  \subfigure[Fraction of compromised nodes ($P_c$)]{
    \includegraphics[width=0.25\textwidth, height=0.19\textwidth]{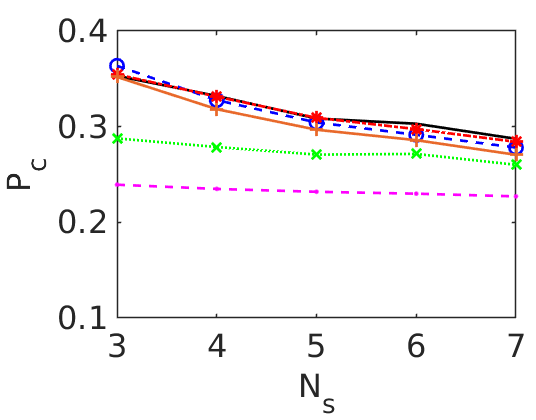}}\hspace{-0.5em}
  \subfigure[Size of a giant component ($S_g$)]{
    \includegraphics[width=0.24\textwidth, height=0.19\textwidth]{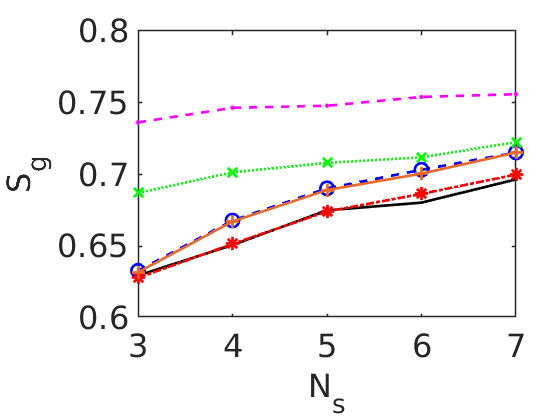}}\hspace{-0.85em}
  \subfigure[Software diversity ($SD$)]{
    \includegraphics[width=0.24\textwidth, height=0.19\textwidth]{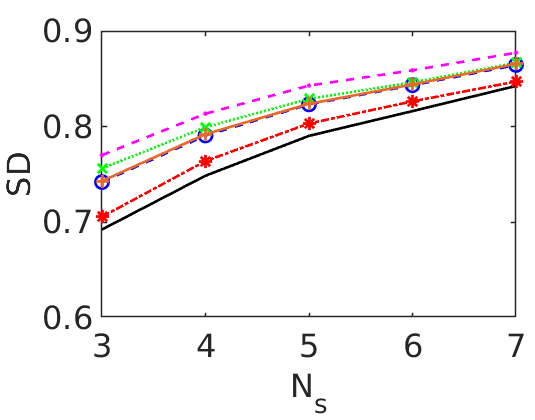}}
    \subfigure[Defense cost ($D_c$)]{
    \includegraphics[width=0.24\textwidth, height=0.19\textwidth]{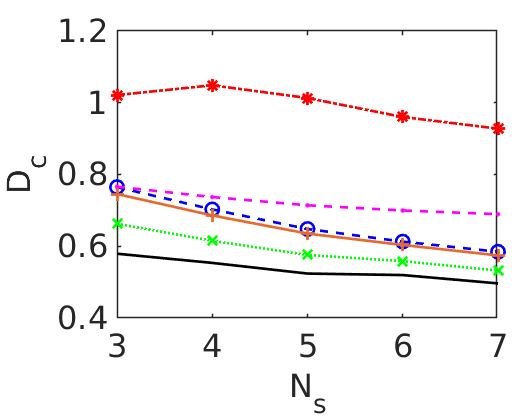}}\hspace{-0.85em}  
    \caption{Effect of the number of software packages ($N_s$) under a medium network.}
\label{fig_n_s_mn}
\end{figure*}
\label{subsec:results_dense}

\subsubsection{Effect of Varying the Fraction of Initial  Attacks ($P_a$)} Fig.~\ref{fig_p_a_dn} shows the effect of varying the attack density ($P_a$) on the performance of the six schemes in terms of the four metrics in Section~\ref{subsec:metrics} under the dense network, whose network topology and degree distribution are shown in Fig. 1 (a) of the appendix file.  
%the following is commented out because it is already discussed earlier in Section 5.3.3
%Under this dense network, $k=1$ and $l=1$ are used to reduce computational complexity as no sensitivity over the metrics in Section~\ref{subsec:metrics} is observed. 
We observe that increasing the percentage of attackers ($P_a$) decreases software diversity ($SD$) and the size of the giant component ($S_g$) while increasing the percentage of compromised nodes ($P_c$) and the defense cost ($D_c$). We note that when more nodes are compromised, the defense cost would also increase since it requires more site percolation based adaptations to be performed when compromised nodes are detected by the IDS (i.e., for disconnecting all edges of a detected, compromised node).

The overall performance order with respect to $P_c$ (representing network security) and $S_g$ (representing network connectivity and resilience) is observed as: SDA with optimal $\rho$ (set at $-0.6$) $\geq$ SDA with $\rho = 0 \geq \text{Random-Graph-C} \approx \text{No-A} \geq$ SDA with $\rho =1 \geq$ Random-A.  It is apparent that the network density of a given network significantly affects both security and performance since SDA with $\rho=-0.6$ and SDA with $\rho=0$ have relatively fewer edges after adaptation and perform better than the other schemes in terms of $P_c$, $S_g$ and $SD$ (i.e., the average software diversity level), as shown in Figs.~\ref{fig_p_a_dn} (a)-(c). 

In Fig.~\ref{fig_p_a_dn} (d), SDA with optimal $\rho$ (set at $-0.6$) also shows significant resilience with relatively low defense cost ($D_c$) as $P_a$ increases.  The overall performance order for other five schemes in $D_c$ is:  Random-Graph-C $\geq$ Random-A $\geq$ SDA with $\rho =1 \geq$ SDA with $\rho =0 \geq$ No-A. Not only that SDA schemes outperform the counterpart baseline schemes in $P_c$, $S_g$, and $SD$, but also the defense cost of SDA schemes are significantly lower than that of Random-Graph-C and are comparable with Random-A (e.g., compared to SDA with optimal $\rho =-0.6$) and No-A (e.g., compared to SDA with $\rho =0$).  This is a significant merit as SDA-based schemes outperform the counterpart baseline schemes with relatively low defense cost.

\subsubsection{Effect of Varying the Number of Software Packages ($N_s$)}

Fig.~\ref{fig_n_s_dn} shows the effect of varying the number of software packages available ($N_s$) on the performance of the six schemes with respect to the metrics defined in Section~\ref{subsec:metrics} under the dense network. We observe that increasing the number of software packages available ($N_s$) increases software diversity ($SD$) and the size of the giant component ($S_g$) while decreasing the percentage of compromised nodes ($P_c$) and the defense cost ($D_c$). Note that based on the concept of $N$-version programming, the number of software packages ($N_s$) here refers to the number of versions being implemented for the same piece of software. Hence, as $N_s$ increases, the software diversity strength increases, resulting in a decrease of the percentage of nodes being compromised due to attacks, an increase of the network connectivity, and a decrease of the defense cost because less nodes are being compromised.

The overall performance order in $P_c$, $S_g$, and $SD$ is very similar to what we observed in Fig.~\ref{fig_p_a_dn}, with SDA with optimal $\rho=-0.6$ 
%and SDA with $\rho=0$ 
outperforming all other schemes. 
%SDA with $\rho=1$ is sensitive to $N_s$ and performs better than No-A when $N_s$ becomes relatively larger. 
For $D_c$, SDA with optimal $\rho=-0.6$ generates a defense cost comparable to that generated by Random-A and in-between those generated by No-A (lowest cost) and Random-Graph-C (highest cost).  
%incur slightly more cost than No-A while generating significantly lower cost than Random-Graph-C.    

\subsection{Comparative Performance Analysis Under a Medium Network}  \label{subsec:results_mn}
%fig 6
\begin{figure*}[!ht]
  \centering
  \subfigure{
    \includegraphics[width=0.6\textwidth, height=0.02\textwidth]{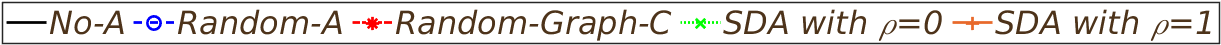}}\vspace{-1em}
   
   \setcounter{subfigure}{0}
  \subfigure[Fraction of compromised nodes ($P_c$)]{
    \includegraphics[width=0.25\textwidth, height=0.19\textwidth]{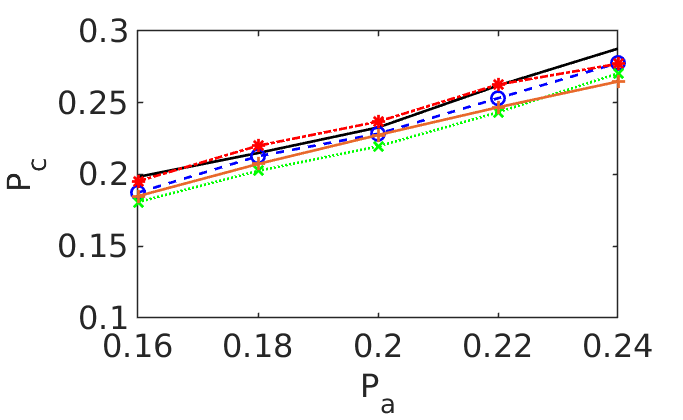}}\hspace{-0.5em}
  \subfigure[Size of a giant component ($S_g$)]{
    \includegraphics[width=0.24\textwidth, height=0.19\textwidth]{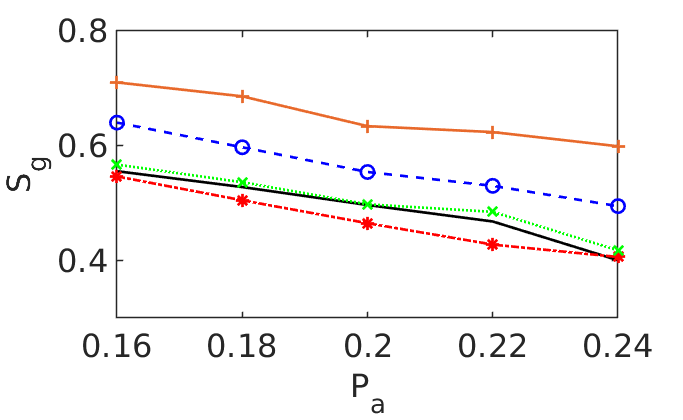}}\hspace{-0.5em}
  \subfigure[Software diversity ($SD$)]{
    \includegraphics[width=0.24\textwidth, height=0.19\textwidth]{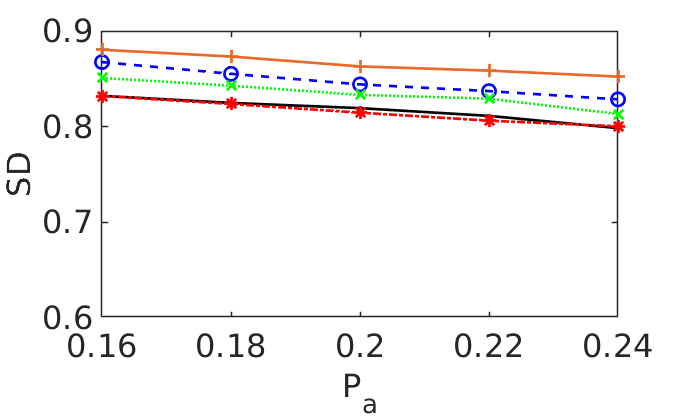}}
  \subfigure[Defense cost ($D_c$)]{
    \includegraphics[width=0.24\textwidth, height=0.19\textwidth]{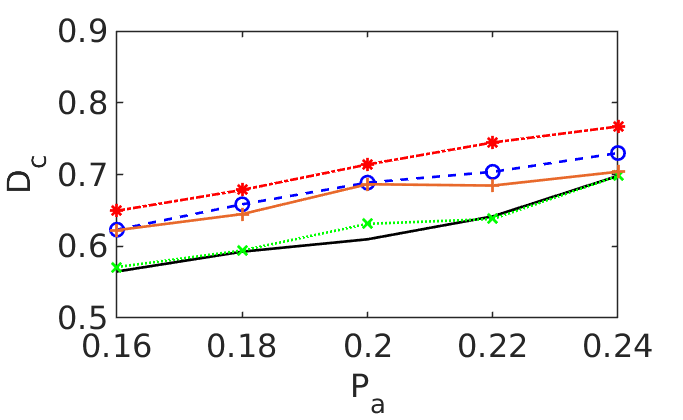}}\hspace{-0.85em}  
    \caption{Effect of varying the fraction of attackers ($P_a$) under a sparse network.}
\label{fig_p_a_sn}
\end{figure*}

%fig 7
\begin{figure*}[!ht]
  \centering
  \subfigure{
    \includegraphics[width=0.6\textwidth, height=0.02\textwidth]{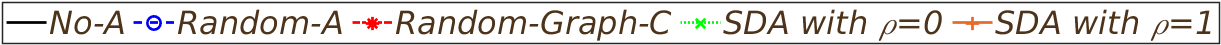}}\vspace{-1em}
   
   \setcounter{subfigure}{0}
  \subfigure[Fraction of compromised nodes ($P_c$)]{
    \includegraphics[width=0.25\textwidth, height=0.19\textwidth]{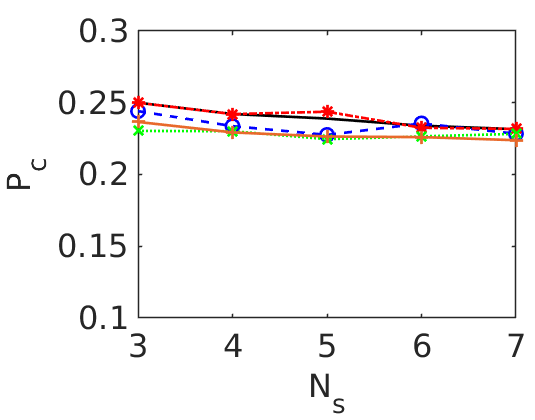}}\hspace{-0.4em}
  \subfigure[Size of a giant component ($S_g$)]{
    \includegraphics[width=0.24\textwidth, height=0.19\textwidth]{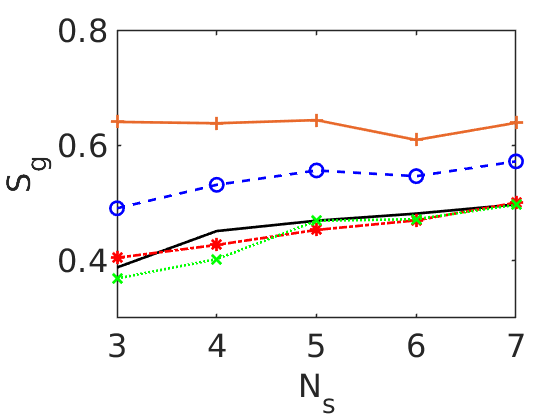}}\hspace{-0.85em}
  \subfigure[Software diversity ($SD$)]{
    \includegraphics[width=0.24\textwidth, height=0.19\textwidth]{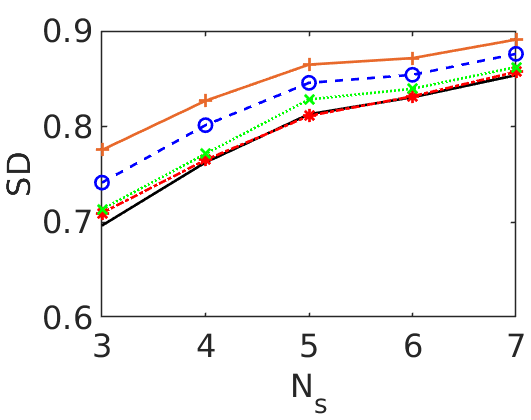}}
  \subfigure[Defense cost ($D_c$)]{
    \includegraphics[width=0.24\textwidth, height=0.19\textwidth]{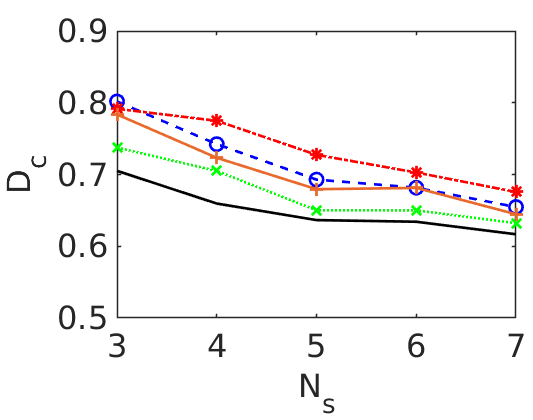}}\hspace{-0.85em}  
    \caption{Effect of the number of software packages ($N_s$) under a sparse network.}
\label{fig_n_s_sn}
\vspace{-3mm}
\end{figure*}
%the following is commented out because it is already discussed earlier in Section 5.3.3
%Similar to the experiment under the dense network, we also observed insensitivity when $k$ and $l$ are varied (see Sections D and E of the appendix file).  Hence, we use $k=l=1$ for high efficiency.

\subsubsection{Effect of Varying the Fraction of Initial  Attacks ($P_a$)}

Fig.~\ref{fig_p_a_mn} demonstrates the effect of varying the percentage of initial attacks on metrics defined in Section~\ref{subsec:metrics} under the medium network, whose network topology and degree distribution are shown in Fig.1 (b) of the appendix file.  Similar to Fig.~\ref{fig_p_a_dn},  Fig.~\ref{fig_p_a_mn} also shows that increasing the percentage of attackers ($P_a$) decreases software diversity ($SD$) and the size of the giant component ($S_g$) while increasing the percentage of compromised nodes ($P_c$) and the defense cost ($D_c$).

The overall performance order in terms of $P_c$ (representing network security) and $S_g$ (representing network connectivity and resilience) is: SDA with optimal $\rho = -0.4 \geq$ SDA with $\rho =0 \geq$ SDA with $\rho =1 \approx$ Random-A $\geq$ Random-Graph-C $\approx$ No-A.  In terms of $SD$ (software diversity), a similar performance order is observed except that Random-Graph-C has a higher $SD$ than No-A. These results demonstrate that SDA schemes clearly are more effective than traditional software shuffling schemes that do not change the network topology (e.g., Random-Graph-C). For the defense cost($D_c$), the overall performance order (the lower cost the better) is: Random-Graph-C $\geq$ SDA with optimal $\rho =-0.4 \geq$ Random-A $\approx$ SDA with $\rho =1 \geq$ SDA with $\rho =0 \geq$ No-A.  Again these results support the claim that SDA-based schemes incur relatively low cost, while outperforming all counterpart baseline schemes in $P_c$, $S_g$, and $SD$.

%\vspace{-2mm}
\subsubsection{Effect of Varying the Number of Software Packages ($N_s$)}  Fig.~\ref{fig_n_s_mn} shows the effect of $N_s$ on performance under the medium network.  We again observe that increasing the number of software packages available ($N_s$) increases software diversity ($SD$) and the size of the giant component ($S_g$) while decreasing the percentage of compromised nodes ($P_c$) and the defense cost ($D_c$). The overall performance order is the same as that in Figs.~\ref{fig_p_a_mn}, with SDA with optimal $\rho=-0.4$ 
outperforming all other schemes in terms of $SD$, $S_g$, and $P_c$ and performing comparably  
to Random-A in terms of $D_c$. By comparing Fig.~\ref{fig_n_s_mn} (for the medium dense network) with Fig.~\ref{fig_n_s_dn} (for the dense network), we also observe that SDA with optimal $\rho$ is more effective in the dense network. We attribute this to node density. That is, SDA is more effective when there are many node connections between nodes in the network allowing SDA to effectively decide which edges to add or remove to effectively maximize software diversity ($SD$) and the size of the giant component ($S_g$) thereby minimizing the percentage of compromised nodes ($P_c$).

%the following is commented since there is no concise information being summarized
%Similar to results in Figs.~\ref{fig_n_s_dn}, SDA with $\rho =0$ and SDA with $\rho =-0.4$ are highly resilient even under a very small $N_s$, showing a steady performance across all the ranges of $N_s$ considered.  On the other hand, the performance of the other counterparts, including Random-A and Random-Graph-C and No-A, increases with higher $N_s$ because more available software packages can significantly contribute to increasing the extent of software diversity.  

% the following is commented out because it is true only for No-A which is not important
%Similar to previous results shown in Figs.~\ref{fig_n_s_dn}, the shuffling cost reaches its maximum with $N_s=4$. 

%{\bf The summary of the key findings in Fig.~\ref{fig_s_er}}: (1) the outperformance of the best schemes (i.e., SDA with $\rho = 0.2$ and SDA $\rho = 0.2$ plus Shuffling) is obvious over the performance of other counterparts; (2) SDA with a sufficiently low threshold (i.e., $\rho = 0.1$) significantly decreases its performance as more software packages are available; and (3) more software packages available lead to more shuffling cost when the shuffling technique is used (i.e., Shuffling Only and SDA $\rho = 0.2$ plus Shuffling).

\subsection{Comparative Performance Analysis Under a Sparse Network} \label{subsec:results_sparse}
%the following is commented out because it is already discussed earlier in Section 5.3.3
%Under the sparse network, we fix $k=2$ and $l=1$ to show the best performance with the minimum computational complexity.  Again the insensitivity across varying $k$ and $l$ is discussed in Sections D and E of the appendix file.

\subsubsection{Effect of Varying the Fraction of Initial  Attacks ($P_a$)}
Fig.~\ref{fig_p_a_sn} shows the effect of varying the initial attack density ($P_a$) on the performance of the five schemes with respect to the 4 performance metrics discussed in Section \ref{subsec:metrics} under the sparse network, whose network topology and degree distribution are shown in Fig. 1 (c) of the appendix file.  
Unlike in the cases of medium and dense networks, the SDA with optimal $\rho$ scheme in the sparse network is the same as the SDA with $\rho=1$ scheme which restores all edges from the lost edges in Step 1 (i.e., $\rho =1$). Therefore, we only show comparative experimental results of the five schemes. 

In the sparse network, the degrees of most nodes are very small, implying that nodes are minimally connected where most nodes only have 1-3 neighbors at most. This means that the network itself is relatively much less vulnerable to epidemic attacks because the attackers inherently cannot reach many nodes to compromise due to network sparsity.  On the other hand, this means that when there is a higher percentage of attackers, the damage upon an attack success (i.e., failing or compromising a node) is more detrimental by resulting in a much smaller size of the giant component representing a significantly lower network resilience (or availability), which introduces a great hindrance to providing continuous services due to a lack of paths available from a source to a destination.
This trend can be clearly observed with the sharp decrease in the size of the giant component ($S_g$) under high attack density (i.e., $P_a=0.24$), when compared to the corresponding results under the medium network (i.e., Fig.~\ref{fig_p_a_mn} (b)). 
A more interesting result is that the overall performance trend does not follow the previous results shown under the dense network (i.e., Fig.~\ref{fig_p_a_dn}) and medium network (i.e., Fig.~\ref{fig_p_a_mn}) which have a sufficiently larger number of edges than the sparse network.
The performance order in $S_g$ is: SDA with optimal $\rho = 1 \geq$ Random-A $\geq$ Random-Graph-C $\approx$ No-A $\geq$ SDA with $\rho =0$.  Since the original network itself is sparsely connected, SDA with $\rho=0$ is not as effective as shown in our previous results for $S_g$ under the dense network (see Fig.~\ref{fig_p_a_dn}) and medium network (see Fig.~\ref{fig_p_a_mn}).  
%We also observed the optimal $\rho$ is $1$, so we did not show the case at the optimal $\rho$ because it is the same as SDA with $\rho =1$, which was one of SDA counterpart baseline schemes.
SDA with optimal $\rho =1$ with all edges restored from the lost edges in Step 1 performs the best in $S_g$.
This result is reasonable because the sparse network does not need to disconnect more edges because it is already sparse enough and significantly less vulnerable to epidemic attacks.

The overall performance with respect to $P_c$ is very similar among all the five schemes, with slightly better results in two SDA schemes. Similarly to the result shown for the dense network, Random-Graph-C exhibits the same level of performance as No-A in $P_c$ and $S_g$, but with a higher software diversity ($SD$).  This indicates the advantage of topology-aware adaptation in a sparse network. For the defense cost ($D_c$) the performance order is: Random-Graph-C $\geq$ Random-A $\geq$ SDA with optimal $\rho =1 \geq$ SDA with $\rho =0 \geq$ No-A.  It is interesting to observe that all SDA-based schemes incur a lower defense cost than Random-A and Random-Graph-C possibly due to fewer compromised nodes in the system and thus less frequent IDS interventions.

%{\bf The summary of the key findings in Fig.~\ref{fig_attack_density_sparse}}: (1) Unlike the previous results in the random network shown in Fig.~\ref{fig_attack_density_er}, the existing counterpart, Graph-C, performs the best among all due to the significantly low network connectivity; (2) the impact of higher attack density under this sparse network is significantly larger than that under the random network; and (3) the shuffling related strategies (i.e., Shuffling Only and SDA with $\rho=0.2$ plus Shuffling) under the sparse network contributes to improving performance better than than those under the random network.

\subsubsection{Effect of Varying the Number of Software Packages ($N_s$)} Fig.~\ref{fig_n_s_sn} shows the effect of varying the number of software packages available ($N_s$) on the performance of the five schemes under the sparse network.  As expected, as $N_s$ increases, $SD$ (software diversity) increases and $D_c$ (defense cost) decreases. As $N_s$ increases, $S_g$ (size of the giant component) also increases for all schemes except for the SDA with optimal $\rho=1$ scheme.   
The reason is that when $\rho=1$, SDA will restore all edges removed in Step 1 (see Step 1 in Algorithm~\ref{algo:SDA}). When $N_s$ is higher, fewer edges will be removed in Step 1 because of a smaller probability that two neighbor nodes will have the same software package. Consequently, when $N_s$ is higher, the very same smaller number of edges will be added back in Step 2 (see Step 2 in Algorithm~\ref{algo:SDA}), thus resulting in the size of the giant component in the shuffled topology not necessarily larger than the one when $N_s$ is lower.

By comparing Fig.~\ref{fig_n_s_sn} (for the sparse network) with Fig.~\ref{fig_n_s_mn} (for the medium dense network) and Fig.~\ref{fig_n_s_dn} (for the dense network), we notice that SDA with optimal $\rho$ is most effective in the dense network. We conclude that our proposed SDA algorithm is most effective in a dense network under which SDA can effectively decide which edges among many to add or remove to effectively maximize software diversity ($SD$) and the size of the giant component ($S_g$) as well as minimizing the percentage of compromised nodes ($P_c$).

%{\bf The summary of the key findings in Fig.~\ref{fig_s_sparse}}: (1) SDA-based strategies do not perform well compared to the results observed under the random network while Graph-C performs the best among all and shuffling based strategies perform fairly well; (2) although Graph-C performs the best in the software diversity and the size of a giant component, it incurs high adaptation cost; (3) shuffling based approaches fairly performs well as a second best in the software diversity and the size of a giant component while incurs much less cost; and (2) the hybrid approach, SDA with $\rho = 0.2$ plus Shuffling, takes the benefits of both SDA and shuffling so it can achieve as much as Shuffling Only in the software diversity and the size of a giant component while incurring a significantly low cost in shuffling.

%Overall we clearly observe the effectiveness of SDA-based schemes over the baseline schemes. In particular, we observe much higher performance when the SDA-based schemes are applied under dense networks than sparse networks.

\section{Conclusions} \label{sec:conclusion}
%\& Future Work
\subsection{Summary}
In this section, we summarize the contributions of this work:
\begin{itemize}
\item We proposed a software diversity metric based on vulnerabilities of attack paths reachable to each node. We called this scheme `software diversity-based adaptation' (SDA) and used it to adapt edges to generate a resilient network topology that can minimize security vulnerability while maximizing network resilience (or connectivity) to provide seamless services under epidemic attacks. 
\item We conducted extensive simulation experiments in order to demonstrate the performance of the proposed SDA scheme compared against other existing counterpart baseline schemes (i.e., random adaptation, random graph coloring, and no adaptation). Via extensive simulation experiments, we found our proposed SDA scheme outperforms counterpart baseline schemes in terms of the fraction of compromised nodes by epidemic attacks, the size of the giant component, and the level of software diversity. In addition, we analyzed the defense cost associated with each scheme and proved the proposed SDA scheme incurs comparable defense cost over existing counterparts.
\item We also identified the optimal setting for executing SDA to meet the imposed performance goals. This allows each node to efficiently compute its software diversity value and use it for adapting edges to maximize its software diversity, leading to minimizing security vulnerability while maximizing network connectivity.
\item We conducted an extensive simulation study with four different real networks in order to investigate the effect of network density on the optimal setting of SDA under which it can best achieve the dual goals of security (i.e., minimum vulnerability) and performance (i.e., maximum network connectivity).
\item We effectively incorporated the techniques of percolation theory in the network science domain into software diversity-based security analysis in the computer science domain.  To be specific, in terms of the computer science perspective, the proposed software diversity metric used attack path vulnerabilities, which are derived based on software vulnerabilities of the intermediate nodes on the attack paths.  On the other hand, in terms of the network science perspective, this work also adopted percolation theory to examine the effect of software diversity-based edge adaptation on network resilience measured by the size of the giant component. Based on the rationale that network interconnectivity can increase both network vulnerability and network connectivity~\cite{Barabasi2016}, this work addressed the tradeoff relationship in the context of cybersecurity, which has not been addressed in the literature.  
\end{itemize}

\subsection{Key Findings}
From our extensive simulation experiments, we obtained the following key findings: 
\begin{itemize}
\item Overall under epidemic attacks, more interconnectivity between nodes in a network introduces higher security vulnerability while bringing a larger size of the giant component, implying higher network connectivity.  In addition, when two nodes use the same software package where the vulnerability of the software package is known to an attacker, it provides a high advantage to the attacker.  How nodes are connected to each other is highly critical in determining the network's vulnerability to epidemic attacks.
\item Even if two network topologies have the same network density (i.e., the same number of edges), how nodes are connected to each other can vastly change the extent of the security vulnerability to epidemic attacks.  It is even possible that a sparser network may introduce more security vulnerability than a denser network depending on how the nodes are connected to each other.
\item It is not necessary to consider the entire network topology for each node to make effective edge adaptation decisions to minimize security vulnerability while maximizing network connectivity.  
%Although our proposed SDA algorithm does not consider many attack paths or longer paths to estimate vulnerability-based software diversity, 
Our SDA algorithm allows each node to make effective decisions on edge adaptation in a lightweight manner. This is because edge adaptation decisions are determined based on ranking of node software vulnerability values, which is more flexible than using a threshold, to achieve the dual goals of security and performance.
%regardless of whether the vulnerability value itself is accurate or not (i.e., considering the upper bound suffices for edge adaptation decision making). 
\item Under medium dense and dense networks, our SDA scheme significantly outperforms existing counterpart baseline schemes. However, under the sparse network, although our SDA scheme still outperforms other schemes, the difference was less significant.  We conclude that our SDA scheme is most effective in a dense network under which SDA can effectively decide which edges among many existing connections to add or remove to effectively maximize software diversity and the size of the giant component as well as minimizing the percentage of compromised nodes.
%This is because in a sparse network, there is not much room to improve the security as the network itself is already too sparse, not allowing the effective propagation of epidemic attacks while achieving a larger size of the giant component is not inherently easy due to the network sparsity.
\item Our proposed SDA scheme is extremely resilient to harsh environments. The performance gain relative to counterpart baseline schemes increases as the environment is harsher, i.e., as the percentage of attackers increases or as the number of the software packages decreases. This proves the high resilience of the proposed SDA scheme under a highly disadvantageous environment.
%\item there exists different optimal network density with respect to network connectivity for different types of networks based on the sensitivity analysis of our SDA schemes
%\item our optimal SDA scheme, based on intelligent network topology adaptation, performs the best among all six schemes considered in this work under three types of networks with different network density
%\item our software diversity metric is more correlated to performance metrics when the network is sparser
\end{itemize}

\begin{comment}
\subsection{Future Work}
We plan to take the following future research directions: 
\begin{itemize}
\item We will consider system dynamics derived from node mobility and node/edge adaptation upon knowing the security status of nodes, which have not been considered in this present work.  We will link our current software diversity-based network adaptation strategy as a diversity-based network topology shuffling technique in moving target defense.
\item We will consider the multiagent deep reinforcement learning (mDRL) algorithm for each node to autonomously determine edge adaptation decisions which can lead to an optimal solution of the robust network topology for maximizing network resilience under attacks.
\item We will consider a more sophisticated attack type, such as advance persistent threat (APT) attacks in which a highly intelligent attacker persistently conducts multi-staged attacks based on the cyber kill chain~\cite{okhravi2013survey}.
\item We will consider targeted attacks based on various types of centrality metrics and investigate their effect on network resilience and system security.
\end{itemize}
\end{comment}

\bibliographystyle{IEEETranSN}
\bibliography{diversity-bib}

\end{document}

\title{Vulnerability-Aware Resilient Networks: Software Diversity-based Network Adaptation}
\author{Qisheng Zhang, Jin-Hee Cho, \IEEEmembership{Senior Member, IEEE,} Terrence J. Moore, \IEEEmembership{Member, IEEE}, Ing-Ray Chen, \IEEEmembership{Member, IEEE}
\IEEEcompsocitemizethanks{\IEEEcompsocthanksitem
Qisheng Zhang, Jin-Hee Cho, and Ing-Ray Chen are with the Department of Computer Science, Virginia Tech, Falls Church, VA, USA. Email: \{qishengz19, jicho, irchen\}@vt.edu.  Terrence J. Moore is with US Army Research Laboratory, Adelphi, MD, USA. Email: terrence.j.moore.civ@mail.mil.}
}
%\tfootnote{add funding information later}

\markboth{IEEE Transactions on Network Service and Management, vol. X, no. X, Month, 2020}{Zhang \MakeLowercase{\textit{et al.}}: Vulnerability-Aware Resilient Networks: Software Diversity-based Network Adaptation}

\maketitle

\appendices

\section{Algorithms}

The following algorithms are referenced from the main paper. They are as follows:
\begin{itemize}
\item {\bf Algorithm~\ref{algo:SDA-step-1}} ($\mathbf{SDBA}$): Remove edges between two nodes with the same software package, which is Step 1 in the SDA algorithm (Algorithm~\ref{algo:SDA-step-1}).
\item {\bf Algorithm~\ref{algo:SDA_pv}} ($\mathbf{GenPV}$): Generate path vulnerabilities. This algorithm returns a vector $\mathbf{PV}$ that contains the maximum attack path vulnerability where each attack path is a disjoint shortest path from node $i$ to a node with the maximum $k-1$ hop distance.
\item {\bf Algorithm~\ref{algo:SDA_set_threshold}} ($\mathbf{SetEAB}$): Set edge adaptations budget.  This algorithm returns $T^{global}$, which is the total number of edges to be adapted based on a given fraction of edges to be adapted, $\rho$, and a vector of the number of edges to be adapted per node, $\mathbf{T}^{local}$.
\item {\bf Algorithm~\ref{algo:SDA_candidate_add}} ($\mathbf{GEAC}$): Generate edge addition candidates. This algorithm returns a vector of edge candidates to be added based on the lowest software diversity reduction.
\item {\bf Algorithm~\ref{algo:SDA_candidate_remove}} ($\mathbf{GERC}$): Generate edge removal candidates. This algorithm returns a vector of edge candidates to be removed based on the highest software diversity increase.
\item {\bf Algorithm~\ref{algo:SDA_rank_and_adapt}} ($\mathbf{AdaptNT}$): Adapt network topology.  Based on the above two algorithms, $\mathbf{GEAC}$ and $\mathbf{GERC}$, $T^{global}$ number of edges are adapted (removed or added) where each node adapts its edge based on $\mathbf{T}^{local}$ based on $\mathbf{SetEAB}$ above.
\item {\bf Algorithm~\ref{algo:random}} ({\bf Random-A}): This algorithm is one of comparable counterpart schemes and compared with the proposed SDA-based schemes.  This algorithm uses the same procedure in Step 1 as the SDA (i.e., $\mathbf{SDBA}$ in Algorithm 1). In Step 2, an edge addition is performed randomly for the number of nodes lost in Step 1 only based on the condition that two nodes use different software packages.
\item {\bf Algorithm~\ref{algo:epidemic-attacks}}: This algorithm describes how the epidemic attacks are modeled in this work.
\end{itemize}

\begin{algorithm}[!th]
\small{
\caption{Software Diversity-based Basic Adaptation ($\mathbf{SDBA}$)}
\label{algo:SDA-step-1}
\begin{algorithmic}[1]
\State{$N \leftarrow$ The total number of nodes in a network}
\State{$\mathbf{DN} \leftarrow$ A vector containing the number of removed edges per node}
\State{$\mathbf{A} \leftarrow$ An adjacency matrix for a given network with element $a_{ij}$ for $i, j = 1, \ldots, N$}
\State{$\mathbf{S} \leftarrow$ A vector of software packages installed over nodes with element $s_i$ for $i = 1, \ldots, N$}
\State{$\mathbf{A}' \leftarrow$ An adjacency matrix after edges are adapted.}

\\
\State{$\mathbf{SDBA} (N, \mathbf{DN}, \mathbf{A}, \mathbf{S})$}
\Comment{Remove edges between two nodes with the same software package.}
\For{$i:=1$ to $N$}
   \For{$j:=1$ to $N$}
        \If{$(a_{ij} > 0) \wedge s_i == s_j$} 
            \State{$a_{ij} = 0$} 
            \State{$a_{ji} = 0$}
            \State{$\mathbf{DN}(i) = \mathbf{DN}(i) + 1$} \State{$\mathbf{DN}(j) = \mathbf{DN}(j) + 1$} 
        \EndIf
   \EndFor
\EndFor
\State{$\mathbf{return}\  \mathbf{A}'$}  
\end{algorithmic}
}
\end{algorithm}

\begin{algorithm}[!th]
\small{
\caption{Generate Path Vulnerabilities ($\mathbf{GenPV}$)}
\label{algo:SDA_pv}
\begin{algorithmic}[1]
\State{$\mathbf{A} \leftarrow$ An adjacency matrix for a given network with element $a_{ij}$ for $i, j = 1, \ldots, N$}
\State{$\mathbf{S} \leftarrow$ A vector of software packages installed over nodes with element $s_i$ for $i = 1, \ldots, N$}

\State{$\mathbf{APV} \leftarrow$ A vector of vulnerabilities of attack paths $apv_{ij}$ from node $i$ to node $j$ where $j = 1, \ldots, 20$} \Comment{For simplicity, consider maximum 20 shortest disjoint attack paths reachable to node $i$}

\State{$k \leftarrow$ A hop distance given in a node's local network}

\State{$\mathbf{PV} \leftarrow$ A vector of maximum path vulnerabilities for all nodes $i$ where $pv_i$ refers to the maximum attack path vulnerability where the path consists of at most $k-1$-hop distance from node $i$} \Comment{Count one hop from a node to any neighboring node and thus $k$ is decremented by 1.}
\State{$\mathbf{PV} =\mathbf{GenPV} (\mathbf{A},\mathbf{APV},\mathbf{S},k)$}
\For{$i:=1$ to $N$}
%\State{Randomly select at most $memory$ length $k-1$ shortest disjoint paths reachable to node $i$}
\State{$pv_i=\max\limits_{j} apv_{ij}^{k-1}$} 
\EndFor
\State{$\mathbf{return}\ \mathbf{PV}$}
\end{algorithmic}
}
\end{algorithm}

\begin{algorithm}[!th]
\small{
\caption{Set Edge Adaptations Budget ($\mathbf{SetEAB}$)}
\label{algo:SDA_set_threshold}
\begin{algorithmic}[1]
\State{$\mathbf{A} \leftarrow$ An adjacency matrix for a given network with element $a_{ij}$ for $i, j = 1, \ldots, N$}

\State{$\mathbf{DN} \leftarrow$ A vector containing the number of removed edges per node}

\State{$\rho \leftarrow$ A threshold referring to the fraction of edges to be adapted}
\State{$\kappa_i \leftarrow$ Node $i$'s degree}
\State{$T^{global} \leftarrow$ The total number of edges to be adapted}
\State{$\mathbf{T}^{local} \leftarrow$ A vector of the number of edges that are allowed to be adapted per node where an element is denoted by $T^{local}_i$ for node $i$ for $i = 1, \ldots, N$}
\\
\State{$\mathbf{T}^{local}, T^{global} = \mathbf{SetEAB} (\mathbf{DN},\mathbf{A},\rho)$}
%\If{$\rho>0$}
\State{$T^{global}= |\rho \frac{sum(\mathbf{DN})}{2}|$}
%\Else
%\State{$T^{global}=-\rho*\frac{sum(\mathbf{A})}{2}$}
%\EndIf
\State{$\kappa=\frac{sum(\mathbf{A})+T^{global}}{N}$} \Comment{An expected average degree after $T^{global}$ number of edges are adapted}
\If{$\rho>0$}
\State{$T^{local}_i \leftarrow \max(0, \kappa - \kappa_i)$} \Comment{$T^{local}_i$ captures the number of edges that can be added}
\State{$N_{HD}=\sum_{i=1}^N \max(0, \kappa_i-\kappa)-T^{global}$}
\Else
\State{$T^{local}_i \leftarrow \max(0, \kappa_i - \kappa)$} \Comment{$T^{local}_i$ captures the number of edges that can be removed}
\State{$N_{HD}=\sum_{i=1}^N \max(0, \kappa-\kappa_i)-T^{global}$}
\EndIf
\Comment{Positive $N_{HD}$ means that more nodes have higher (or lower) degrees than the expected average degree while negative $N_{HD}$ implies fewer nodes have higher (lower) degrees than the average expected degree after edge adaptations.  Hence, when $N_{HD}$ is positive, edge adaptations should be restricted further.}
\While{$N_{HD}>0$}
\For{$i:=1$ to $N$}
\If{$T^{local}_i>0 \wedge N_{HD}>0$}
\State{$T^{local}_i=T^{local}_i-1$}
\State{$N_{HD}=N_{HD}-1$}
\EndIf
\EndFor
\EndWhile
\State{$\mathbf{return}\ T^{global}, \mathbf{T}^{local}$}
\end{algorithmic}
}
\end{algorithm}

\begin{algorithm}[!th]
\small{
\caption{Generate Edge Addition Candidates ($\mathbf{GEAC}$)}
\label{algo:SDA_candidate_add}
\begin{algorithmic}[1]
\State{$N \leftarrow$ The total number of nodes in a network}
\State{$\mathbf{A} \leftarrow$ An adjacency matrix for a given network with element $a_{ij}$ for $i, j = 1, \ldots, N$}
\State{$\mathbf{A^*}\leftarrow \mathbf{(A+I)}^{2k}$ where $a^*_{ij}$ is 1 when nodes $i$ and $j$ belong to each other's local network or its neighbor's local networks; 0 otherwise.}
\State{$\mathbf{SD} \leftarrow$ A vector of software diversity values, $sd_i$ for all nodes $i = 1, \ldots, N$}
\State{$\mathbf{SV} \leftarrow$ A vector of the vulnerabilities associated with software packages}
\State{$\mathbf{S} \leftarrow$ A vector of software packages installed over nodes with element $s_i$ for $i = 1, \ldots, N$}
\State{$\mathbf{PV} \leftarrow \mathbf{GenPV} (\mathbf{A},\mathbf{APV},\mathbf{S},k)$ }
\State{$\mathbf{DN} \leftarrow$ A vector containing the number of removed edges per node}
\State{$\mathbf{T}_{local} \leftarrow \mathbf{GetEdgesToAdapt} (\mathbf{DN}, \mathbf{A}, \rho)$} \\
\State{$\mathbf{\mathbf{add\_candidate}}$ = $\mathbf{GEAC}$($\mathbf{A},\mathbf{SD},\mathbf{SV},\mathbf{S},\mathbf{PV}, \mathbf{T}^{local}$)}
\State{$r \leftarrow$ counter initialized at 0.}
    \For{$i:=1$ to $N$}
    \For{$j:=i+1$ to $N$}
    \If{$a^*_{ij}>0 \wedge a_{ij}==0 \wedge s_i\neq s_j$}
    \State{$\mathbf{sd\_diff\_sum}(r)= (sd_i - sd'_i) +(sd_j - sd'_j)$} \\
    \Comment{Sum of the improved software diversity values by both node $i$ and node $j$ where the expected software diversity values of nodes $i$ and $j$ after edge addition adaptations are $sd'_i = sd_i (1-sv_{s_i} pv_j), sd'_j = sd_j (1-sv_{s_j} pv_i)$ based on Eq.~(3) of the main paper (i.e., software diversity metric)}
    \State{$r = r+1$}
%\If{$count_i<T^{local}_i \wedge count_j<T^{local}_j$}
    %\State{Add tuple $(i,j,temp)$ to $\mathbf{candidate}$}
    %\State{$max_i=\max(max_i,temp)$}
    %\State{$max_j=\max(max_j,temp)$}
    %\State{$count_i=count_i+1;count_j=count_j+1$}
    %\Else
    %\If{$temp<\min(max_i,max_j)$}
    %\State{Add tuple $(i,j,temp)$ to $\mathbf{candidate}$}
    %\State{$max_i=max_j=temp$}
    %\EndIf
    %\EndIf
    \EndIf
\EndFor
\EndFor
\State{Rank $\mathbf{sd\_diff\_sum}$ in an ascending order and capture top $\mathbf{T}^{local}$ edges in $\mathbf{add\_candidate}$}
\State{$\mathbf{return}\  \mathbf{add\_candidate}$}  
\end{algorithmic}
}
\end{algorithm}

\begin{algorithm}[!th]
\small{
\caption{Generate Edge Removal Candidates ($\mathbf{GERC}$)}
\label{algo:SDA_candidate_remove}
\begin{algorithmic}[1]
\State{$N \leftarrow$ The total number of nodes in a network}
\State{$\mathbf{A} \leftarrow$ An adjacency matrix for a given network with element $a_{ij}$ for $i, j = 1, \ldots, N$}
\State{$\mathbf{A^*}\leftarrow \mathbf{(A+I)}^{2k}$ where $a^*_{ij}$ is 1 when nodes $i$ and $j$ belong to each other's local network or its neighbor's local networks; 0 otherwise.}
\State{$\mathbf{SD} \leftarrow$ A vector of software diversity values, $sd_i$ for all nodes $i = 1, \ldots, N$}
\State{$\mathbf{SV} \leftarrow$ A vector of the vulnerabilities associated with software packages}
\State{$\mathbf{S} \leftarrow$ A vector of software packages installed over nodes with element $s_i$ for $i = 1, \ldots, N$}
\State{$\mathbf{PV} \leftarrow \mathbf{GenPV} (\mathbf{A},\mathbf{APV},\mathbf{S},k)$ }
\State{$\mathbf{DN} \leftarrow$ A vector containing the number of removed edges per node}
\State{$\mathbf{T}_{local} \leftarrow \mathbf{GetEdgesToAdapt} (\mathbf{DN}, \mathbf{A}, \rho)$} \\
\State{$\mathbf{remove\_candidate}$ = $\mathbf{GERC}(\mathbf{A},\mathbf{SD},\mathbf{SV},\mathbf{S},\mathbf{PV},\mathbf{T}^{local}$)}
%\If{$a_{ij}==1$}
%\State{$temp=sd_i (\frac{1}{1-sv_{s_i} pv_j}-1)+sd_j (\frac{1}{1-sv_{s_j} pv_i}-1)$} 

    \For{$i:=1$ to $N$}
    \For{$j:=i+1$ to $N$}
    \If{$a_{ij}>0$}
    \State{$\mathbf{sd\_diff\_sum}(r)= (sd'_i - sd_i) +(sd'_j - sd_j)$} \\
    \Comment{Sum of the improved software diversity values by both node $i$ and node $j$ where the expected software diversity values of nodes $i$ and $j$ after edge removal adaptations are $sd'_i =  sd_i/(1-sv_{s_i} pv_j), sd'_j = sd_j/(1-sv_{s_j} pv_i)$ based on Eq.~(3) of the main paper (i.e., software diversity metric)}
    \State{$r = r+1$}
\EndIf
\EndFor
\EndFor
\State{Rank $\mathbf{sd\_diff\_sum}$ in an descending order and capture top $\mathbf{T}^{local}$ edges in $\mathbf{remove\_candidate}$}
\State{$\mathbf{return}\  \mathbf{remove\_candidate}$}  

%\Comment{estimated SD difference}
%\If{$count_i<T^{local}_i \wedge count_j<T^{local}_j$}
%\State{Add tuple $(i,j,temp)$ to $\mathbf{candidate}$}
%\State{$min_i=\min(min_i,temp)$}
%\State{$min_j=\min(min_j,temp)$}
%\State{$count_i=count_i+1;count_j=count_j+1$}
%\Else
%\If{$temp>\max(min_i,min_j)$}
%\State{Add tuple $(i,j,temp)$ to $\mathbf{candidate}$}
%\State{$min_i=min_j=temp$}
%\EndIf
%\EndIf
%\EndIf
%\State{$\mathbf{return}\ %\mathbf{candidate}$}  
\end{algorithmic}
}
\end{algorithm}

\begin{algorithm}[!th]
\small{
%\caption{Ranking Edges Candidates and Adapting ($\mathbf{RankAndAdapt}$)}
\caption{Adapt Network Topology ($\mathbf{AdaptNT}$)}
\label{algo:SDA_rank_and_adapt}
\Comment{Adapt edges based on Algorithms~\ref{algo:SDA_candidate_add} and~\ref{algo:SDA_candidate_remove}}
\begin{algorithmic}[1]
\State{$N\leftarrow\#$ of nodes in a given network}
\State{$\mathbf{DN} \leftarrow$ a vector with \# of disconnected edges per node}
\State{$\mathbf{A} \leftarrow$ an adjacency matrix for a given network} 
\State{$\mathbf{SV} \leftarrow$ a constant vulnerability vector for software packages}
\State{$\mathbf{S} \leftarrow$ a software package vector for a given network}
\State{$k \leftarrow$ hop distance of local networks}
\State{$l \leftarrow$ search threshold of \# of attack paths}
\State{$\rho \leftarrow$ threshold of the fraction of edges adapted}
\State{$\mathbf{candidate} \leftarrow$ A vector of edge candidates, either $\mathbf{add\_candidate}$ in Algorithms~\ref{algo:SDA_candidate_add} or~\ref{algo:SDA_candidate_remove}}
\State{$\mathbf{T}^{local}, T^{global} = \mathbf{SetEAB} (\mathbf{DN},\mathbf{A},\rho)$ based on Algorithm~\ref{algo:SDA_set_threshold}}
\State{$\mathbf{A}' \leftarrow$ An adjacency network after edges are adapted.}
\\
\State{$\mathbf{A}' = \mathbf{AdaptNT} (\mathbf{A},\mathbf{candidate},\mathbf{T}^{local},T^{global},\rho)$}
%\If{$\rho>0$}
%\State{Sort $\mathbf{candidate}$ $(i,j,temp)$ w.r.t $temp$ in ascend order}
%\Else
%\State{Sort $\mathbf{candidate}$ $(i,j,temp)$ w.r.t $temp$ in descend order}
%\EndIf
\For{$(i,j, \mathbf{sd\_diff\_sum})$ in $\mathbf{candidate}$}
\If{$T^{local}_i T^{local}_j > 0$}
\If{$\rho>0$}
\State{$a_{ij}=a_{ji}=1$}
\Else
\State{$a_{ij}=a_{ji}=0$}
\EndIf
\State{$T^{local}_i=T^{local}_i-1$}
\State{$T^{local}_j=T^{local}_j-1$}
\State{$T^{global}=T^{global}-1$}
\EndIf
\EndFor
\For{$(i,j, \mathbf{sd\_diff\_sum})$ in $\mathbf{candidate}$}
\If{$T^{global}>0$}
\If{$\rho>0 \wedge a_{ij}==0$}
\State{$a_{ij}=a_{ji}=1$}
\State{$T^{global}=T^{global}-1$}
\Else
\If{$\rho<0 \wedge a_{ij}>0$}
\State{$a_{ij}=a_{ji}=0$}
\State{$T^{global}=T^{global}-1$}
\EndIf
\EndIf
\EndIf
\EndFor
\State{$\mathbf{return}\  \mathbf{A'}$} \end{algorithmic}}
\end{algorithm}

\begin{algorithm}[!th]
\small{
\caption{Random Adaptation (Random-A)}
\label{algo:random}
\begin{algorithmic}[1]
\State{$N \leftarrow$ The total number of nodes in a network}
\State{$\mathbf{DN} \leftarrow$ a vector with \# of disconnected edges per node}
\State{$\mathbf{A} \leftarrow$ an adjacency matrix for a given network} 
\State{$\mathbf{S} \leftarrow$ A vector of software packages installed over nodes with element $s_i$ for $i = 1, \ldots, N$}
\State{$\mathbf{SV} \leftarrow$ A vector of the vulnerabilities associated with software packages}
\State{$k \leftarrow$ A hop distance given in a node's local network}
\State{$l \leftarrow$ A maximum number of attack paths considered for estimating a node's software diversity}
\State{$\rho \leftarrow$ A threshold referring to the fraction of edges to be adapted}
\State{$\mathbf{A'} \leftarrow$ An adjacency matrix after edges are adapted.}\\
\State{$\mathbf{A'}=\mathbf{Random}$-$\mathbf{A}(N, \mathbf{DN}, \mathbf{A}, \mathbf{S}, \mathbf{SV}, k, l, \rho)$}
\State{{\bf Step 1:} $\mathbf{A'} = \mathbf{SDBA} (N, \mathbf{DN}, \mathbf{A}, \mathbf{S}, \mathbf{SV}, k, l, \rho)$} \Comment{Remove edges between two nodes with the same software package (see Algorithm~\ref{algo:SDA-step-1}).  $\mathbf{DN}$ is a vector of edges that are removed in Step 1.}
\\
\State{{\bf Step 2:} Add random edges between two disconnected nodes if their software packages are different}
\For{$i:=1$ to $N$}
   \State{$\mathbf{candidate} \leftarrow$ a vector of nodes that can be connected with node $i$ where $a_{ij}==0 \wedge (na_i \cdot na_j > 0)$ for node $j$}
   \State{$\mathbf{visited} \leftarrow$ a vector with $length(\mathbf{candidate})$}
   \For{$j:=1$ to $\mathbf{DN}(i)$} %\Comment{$\mathbf{DN}(i)$ is \# of edges that node $i$ lost in Step 1}
       \State{$r \leftarrow$ A random integer selected from $[0, length(\mathbf{candidate})]$ at random.}       
       \If{$\mathbf{visited}(r)==0 \wedge \mathbf{DN}(r)>0$}
          \State{$a_{ir} = 1$}
          \State{$a_{ri} = 1$}
          \State{$\mathbf{DN}(i)=\mathbf{DN}(i)-1$}
          \State{$\mathbf{DN}(r)=\mathbf{DN}(r)-1$}
          \State{$\mathbf{visited}(r)=1$}
       \Else
          \State{$j=j-1$}
       \EndIf
       \If{$\mathrm{sum}(\mathbf{visited}) == \mathrm{length}(\mathbf{visited})$} \\
       \Comment{all candidate nodes are selected}
          \State{$break$}
      \EndIf
   \EndFor
\EndFor
\State{$\mathbf{return}\  \mathbf{A'}$} \end{algorithmic}	
}
\end{algorithm}

\begin{algorithm}[th!]
\small{
\caption{Epidemic Attacks}
\label{algo:epidemic-attacks}
\begin{algorithmic}[1]
\State{{\bf Input:}} 
\State{$\mathbf{A}\leftarrow$ an adjacency matrix} 
\State{$\mathbf{\sigma}_i\leftarrow$ attacker $i$' vector of exploitable software packages}
\State{$N_s\leftarrow$ a number of software packages available}
\State{$\gamma\leftarrow$ an intrusion detection probability}
\State{$\mathbf{node} \leftarrow$} nodes' attributes, defined in Eq.~(1) of the main paper.
\State{$\mathbf{S} \leftarrow$ a software package vector for a given network}
\State{$\mathbf{SV} \leftarrow$ A vector of the vulnerabilities associated with software packages}
\Procedure{performEpidemicAttacks}{$\mathbf{A}$, $\mathbf{node}$, $\mathbf{SV}$, $\sigma_j$, $N_s$, $\gamma$} 
\State{$spreadDone$ $\leftarrow 0$}
\State{$\mathbf{spread}$: a list with length $N$, initialized at 0} \Comment{To check if node $i$ attempted to compromise its direct neighbors}
  \While{$spreadDone == 0$}
    \For{$i:=1$ to $N$} \Comment{check if $i$ is an attacker}
      \If{$na_i>0$}\Comment{If $i$ is an active attacker}
        \State{$r_1 \leftarrow$ a random real number in $[0, 1]$ based on uniform distribution}
      \If{$nc_i>0$} 
        \If{$r_1 > \gamma\wedge\mathbf{spread}(i)<2$}
        \State{$\mathbf{spread}(i) = \mathbf{spread}(i)+1$}
           \For{$j:=1$ to $N$}
             \If{$a_{ij}> 0 \; \wedge na_j > 0 \wedge nc_j == 0$} \Comment{if $j$ is susceptible}
               \If{$\mathbf{\sigma}_i$ includes $s_j$} \Comment{$i$ knows $s_j$'s vulnerability}
                  \State{$nc_j=1$}  \Comment{$j$ is compromised by $i$}
               \Else
                   \State{$r_2 \leftarrow$ a random real number in $[0, 1]$}
                   \State{$d \leftarrow sv_{s_j}$} %\Comment{$v_{int}$ is the probability that two nodes have an interdependent vulnerability in their software packages installed}
                   \If{$r_2 < d$}
                      \State{$nc_j=1$}  \Comment{$j$ is compromised by $i$}
                      \State{$\sigma_i(\comg{s_j})=1$} \Comment{$i$ learned $s_j$'s vulnerability}
                   \EndIf
               \EndIf
                
             \EndIf
           \EndFor
           \Else
              \State{$na_i=0$} \Comment{$i$ is detected and deactivated for infecting behavior}
              \State{$a_{ij}=0, a_{ji}=0$} \Comment{disconnecting all edges connected to $i$}
         \EndIf %if gamma
        \Else
            \If{$r_1 > \gamma$}
            \State{$na_i=0$} 
            \State{$a_{ij}=0, a_{ji}=0$} 
            \EndIf
              \EndIf
              \EndIf
       \For{$k:=1$ to $N$}
       \If{$na_i > 0 \wedge nc_i > 0 \wedge \mathbf{spread}(k)< 2$} \Comment{each node has two chances to compromise each of its direct neighbors}
          \State{$spreadDone=0$}
          \State{$break$}
       \Else
          \State{$spreadDone=1$}
       \EndIf
    \EndFor
    \EndFor
  \EndWhile
\EndProcedure
\end{algorithmic}	
}
\end{algorithm}

\newpage

\section{Real Network Topologies Used} \label{sec:network-topologies}

This work develops a topology-aware notion of software diversity and, hence it makes sense to study the effect of different network topologies. To meet this objective, we considered three different network datasets, each with different levels of network density. 

\begin{figure*}[!htb]
  \centering
  \subfigure[Dense network (DN) with 1033 nodes and 26747 edges]{
    \includegraphics[width=0.3\textwidth, height=0.2\textwidth]{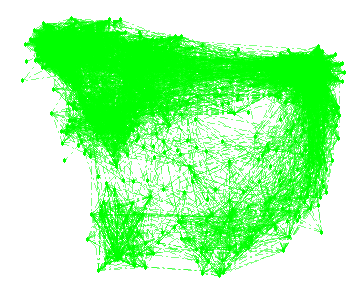}}
  \subfigure[Medium dense network (MN) with 985 nodes and 7994 edges]{
    \includegraphics[width=0.3\textwidth, height=0.2\textwidth]{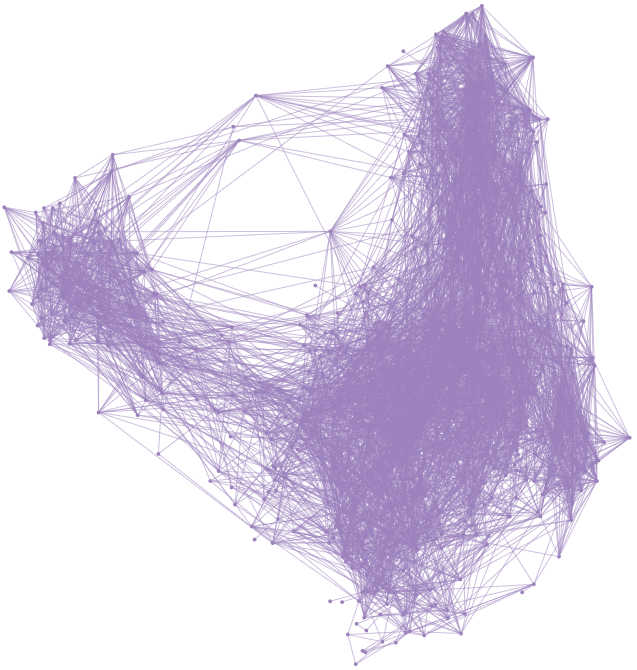}}  
  \subfigure[Sparse network (SN) with 1476 nodes and 3254 edges]{
    \includegraphics[width=0.3\textwidth, height=0.2\textwidth]{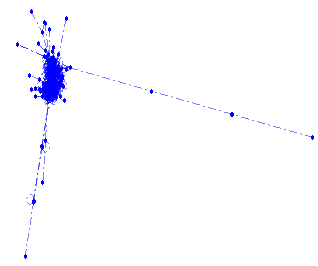}}
  \subfigure[DN's degree distribution]{
    \includegraphics[width=0.3\textwidth, height=0.2\textwidth]{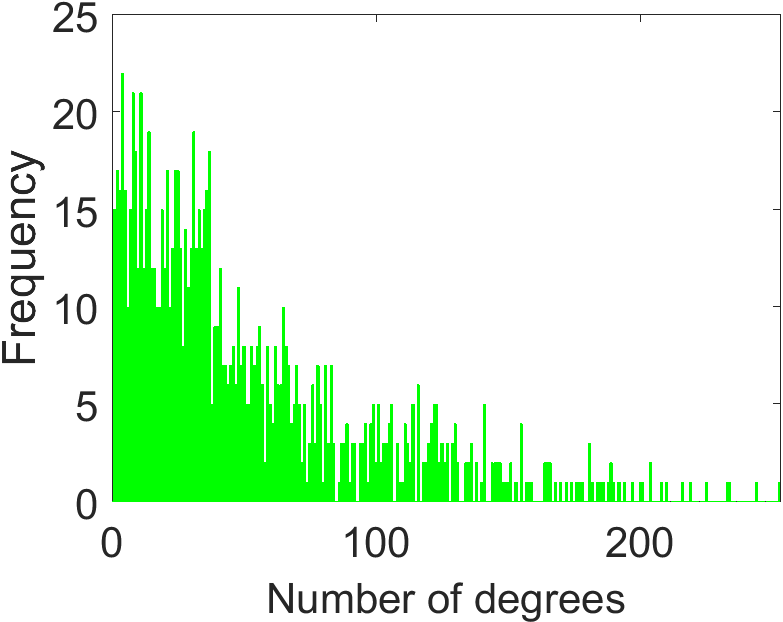}}
  \subfigure[MN's degree distribution]{
    \includegraphics[width=0.3\textwidth, height=0.2\textwidth]{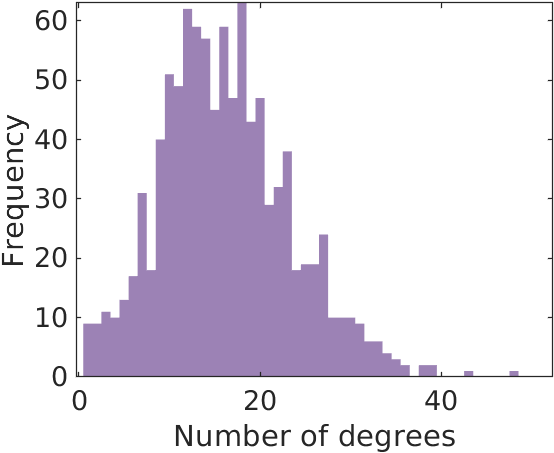}}
  \subfigure[SN's degree distribution]{
    \includegraphics[width=0.3\textwidth, height=0.2\textwidth]{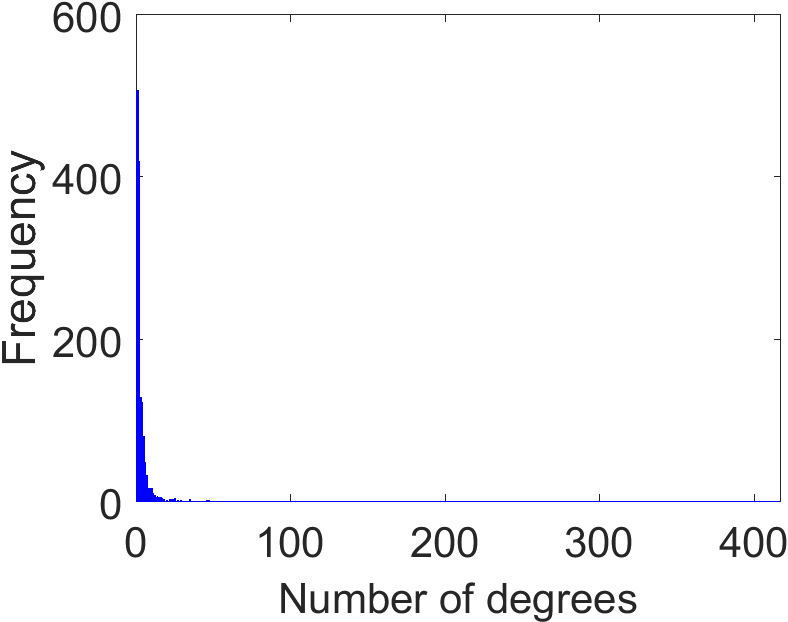}}
\caption{The network topologies used for our experimental study and their degree distributions.}
\label{fig_network_topology}
\end{figure*}

A dense network (DN) is generated from a Facebook ego network, which is a connected component from the one of the 10 ego networks provided by SNAP~\cite{snapnets}. This network has density of approximately $0.05$ with a mean degree near $52$. A visualization of this network is shown in Fig.~\ref{fig_network_topology} (a) and its degree distribution is shown in Fig.~\ref{fig_network_topology} (d).  The distribution is right-skewed despite the high mean.

A medium dense network (MN) is generated from a subset of the Enron email dataset on SNAP~\cite{snapnets}. This medium dense network is generated by the largest connected component of the induced subgraph of nodes ranked between 501 and 1500 by degree. This was done to generate a network with a similar size to the dense network from an original network that starts with less density. The density is $0.016$ with a mean degree near $16$. A visualization of this network is shown in Fig.~\ref{fig_network_topology} (b) and its degree distribution is shown in Fig.~\ref{fig_network_topology} (e).  The distribution has the shape close to a binomial distribution.

A sparse network (SN) is generated from an observation of the Internet at the autonomous systems level~\cite{snapnets}. Edges are removed from the recorded data to ensure a simple network. The density of this network is approximately $0.003$ and the mean node degree is approximately $4.4$. A visualization of this network is shown in Fig.~\ref{fig_network_topology} (c) and its degree distribution is shown in Fig. \ref{fig_network_topology} (f). The distribution skews right and is close to a power-law shape.

%fig1

\section{Experiments Under a Random Network}

The work also considers a random network topology generated by an Erd{\"o}s-R{\'e}nyi (ER) random graph model $G(n,p)$ where $n$ is the number of nodes and $p$ is the connection probability between any pair of nodes~\cite{Newman10}.  This model is ideal to study the effect of network density in a general manner by varying the connection probability $p$ since the network has density approximately $p$ for large $n$. It is used here for a sensitivity analysis and comparative analysis for the methods considered in the paper.
%fig2
\begin{figure*}[!ht]
  \centering
  \subfigure[Erd{\"o}s-R{\'e}nyi random network (ER) with 1000 nodes and 12473 edges using $p=0.025$]{
    \includegraphics[width=0.3\textwidth, height=0.2\textwidth]{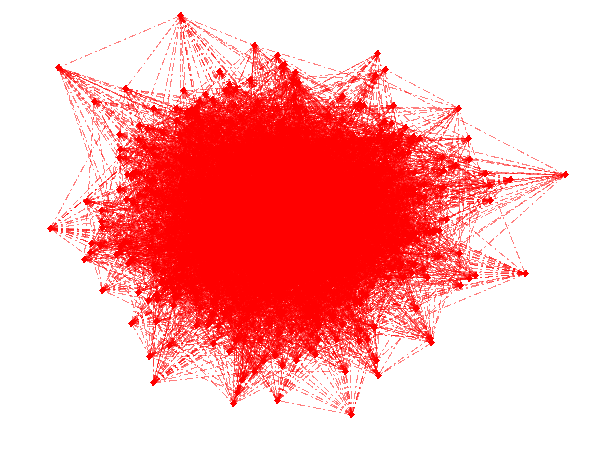}}
  \subfigure[ER's degree distribution]{
    \includegraphics[width=0.3\textwidth, height=0.2\textwidth]{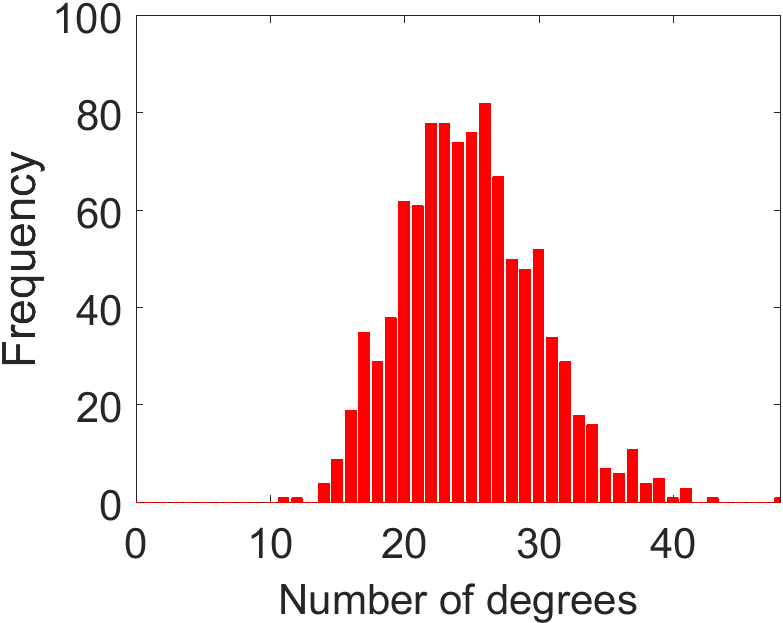}}
\caption{The network topology of a random network used for our experimental study and its degree distribution.}
\label{fig_er_topology}
\end{figure*}

\subsection{Visualization of an ER Network}
A visualization of the example ER network for $n=1000$ and $p=0.025$ is shown in Fig.~\ref{fig_er_topology} (a). This realization of the ER model has density $0.02497$, which is very close to $p$, and has the mean node degree $24.946$, which is close to the expected value $p(n-1)$. The degree distribution of this network realization, shown in Fig.~\ref{fig_er_topology} (b), has is very close to a binomial distribution.

\subsection{Identification of an Optimal Fraction of Edges to be Adapted ($\rho$)}

\begin{figure}[!ht]
  \centering
  \subfigure[Random Network]{
    \includegraphics[width=0.3\textwidth, height=0.22\textwidth]{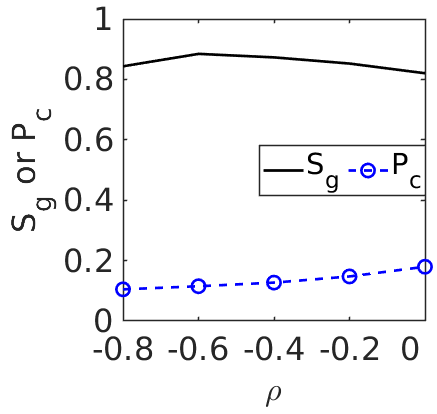}}
    \caption{Effect of the threshold of the fraction of edges adapted ($\rho$) on network connectivity ($S_g$) and security vulnerability ($P_c$).}
\label{fig_er_sensitivity}
\end{figure}

We conducted a sensitivity analysis of the effect of the fraction of edges to be adapted ($\rho$) for the random network here. Fig.~\ref{fig_er_sensitivity} shows the size of the giant component ($S_g$) and the fraction of compromised nodes ($P_c$) versus varying $\rho$.  We used $P_a=0.1$, $N_s=5$, and $p=0.025$ for corresponding simulations. We observe that $S_g$ attains its maximum when $\rho=-0.6$, which will be used as the optimal $\rho$ for the comparative analysis conducted in Section~\ref{Comparative Analysis}.
%fig3

\subsection{Comparative Performance Analysis}
\label{Comparative Analysis}
This section details our comparative analysis of the six schemes introduced in Section 5.2 under a random network. 

\subsubsection{Effect of Network Density Under a Random Network} 
\begin{figure*}[!ht]
  \centering
  \subfigure{
    \includegraphics[width=0.65\textwidth, height=0.02\textwidth]{figs/fig2/legend.png}}
       \setcounter{subfigure}{0}

  \subfigure[Fraction of compromised nodes ($P_c$)]{
    \includegraphics[width=0.26\textwidth, height=0.19\textwidth]{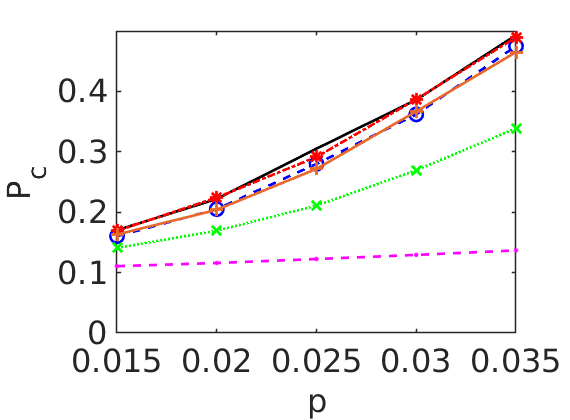}}
  \subfigure[Size of the giant component ($S_g$)]{
    \includegraphics[width=0.24\textwidth, height=0.19\textwidth]{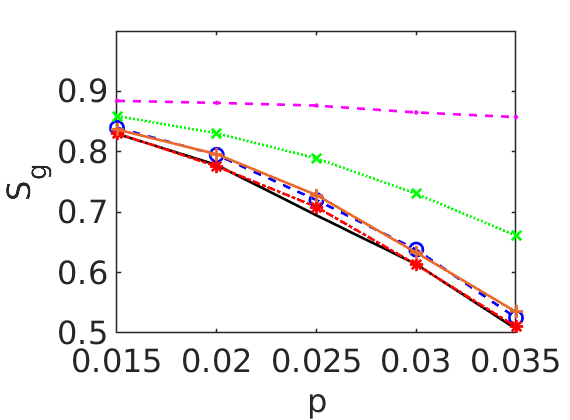}}
  \subfigure[Software diversity ($SD$)]{
    \includegraphics[width=0.24\textwidth, height=0.19\textwidth]{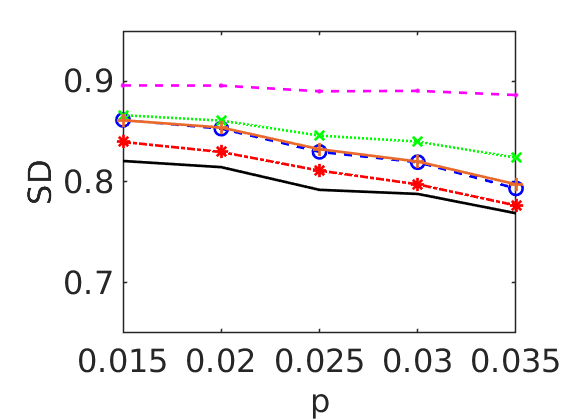}}\hspace{-0.85em} 
  \subfigure[Defense cost ($D_c$)]{
    \includegraphics[width=0.24\textwidth, height=0.19\textwidth]{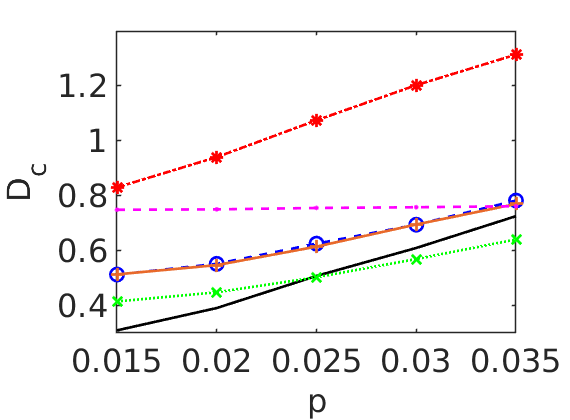}}\hspace{-0.85em}     
    \caption{Effect of the connection probability between two nodes ($p$) under random networks.}
     \label{fig_p_er}
\end{figure*}

Fig.~\ref{fig_p_er} shows how the network density, controlled by the ER connectivity parameter $p$, affects the performance of the six schemes with respect to the four performance metrics listed in Section 5.1 of the main paper. Overall, as the node connection probability $p$ increases, the network is exposed to higher vulnerability with respect to epidemic attacks because the attackers have more neighbors and, hence, more potential nodes to compromise. 
Since the number of software packages available $N_s$ is limited to 5, as density increases, an attacker has a better chance to come into contact with a neighboring node that has a software package shared or previously learned by the attacker. 
Since higher network connectivity naturally leads to higher vulnerability to epidemic attacks, as more nodes are connected, the attacks exhibit a greater impact. After attacks on the original or adapted network, the resulting outcome is a less connected network with a smaller size of the giant component.
This post-attack network also has a lower software diversity value as a fewer number of nodes will be considered in the attack path to estimate the software diversity, as shown in Eq. (3) in Section 4.1. 
Therefore, when $p$ is low, significant performance differences by all schemes are not observed. On the other hand, as $p$ increases, schemes with lower adapted network density (i.e., SDA with $\rho = 0$ and SDA with optimal $\rho = -0.6$) outperform the other counterparts (i.e., No-A, Random-A, Random-Graph-C and SDA with $\rho = 1$). 

%fig6

Overall the best performing method with respect to the three metrics is shown with SDA with $\rho = -0.6$. SDA with $\rho = -0.6$ also shows significant resilience among all four performance metrics. The performance order in the fraction of compromised nodes ($P_c$), the size of the giant component ($S_g$) and software diversity ($SD$) is: SDA with $\rho = -0.6 \geq$ SDA with $\rho = 0 \geq$ SDA with $\rho = 1 \approx$ Random-A $\geq$ Random-Graph-C $\approx$ No-A.  Although Random-Graph-C exhibits slightly better performance than No-A, both schemes perform nearly identically. 
This implies that the contribution of shuffling is minimal under the random network model. 
With respect to the defense cost ($D_c$), the overall performance order follows: SDA with $\rho = 0$ $\approx$ No-A $\geq$ SDA with $\rho = 1$ $\approx$ Random-A $\geq$ SDA with $\rho = -0.6 \geq$ Random-Graph-C. Random-Graph-C incurs the highest cost since the shuffling cost and the cost caused by the IDS are combined when calculating the defense cost using Eq.~(11) in Section 5.1 of the main paper. Although adaptation schemes incur higher cost than No-A, as nodes are more connected with higher $p$, SDA with $\rho = 0$ generates significantly lower cost than the other existing counterparts. 
This is because well-adapted network topologies are less vulnerable to epidemic attacks. In these networks, a significantly fewer number of nodes become compromised and, hence, detected by the IDS, which disconnects all the edges from the detected and compromised nodes.
On the other hand, non-adapted or poorly adapted network topologies result in more actions by the IDS, which will disconnect all the edges of nodes that are detected as compromised to protect the network itself.

\subsubsection{Effect of the Fraction of Initial Seeding Attackers ($P_a$) under a Random Network}

\begin{figure*}[!ht]
  \centering
  \subfigure{
    \includegraphics[width=0.65\textwidth, height=0.02\textwidth]{figs/fig2/legend.png}}\vspace{-1em}
   
   \setcounter{subfigure}{0}
  \subfigure[Fraction of compromised nodes ($P_c$)]{
    \includegraphics[width=0.25\textwidth, height=0.19\textwidth]{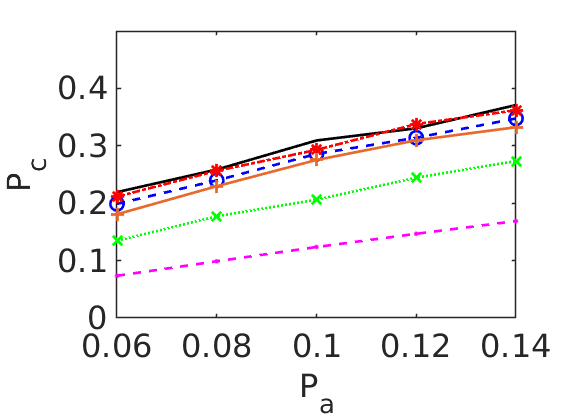}}
  \subfigure[Size of the giant component ($S_g$)]{
    \includegraphics[width=0.24\textwidth, height=0.19\textwidth]{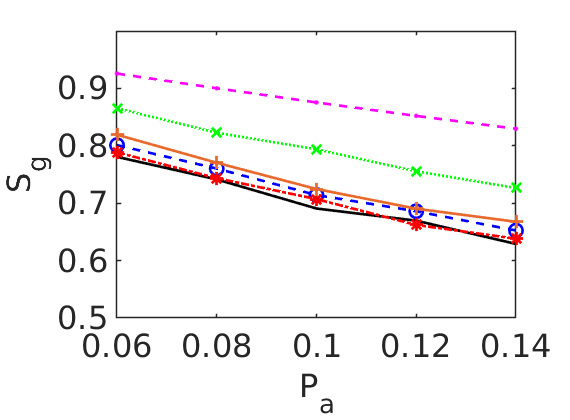}}\hspace{-0.85em}
  \subfigure[Software diversity ($SD$)]{
    \includegraphics[width=0.24\textwidth, height=0.19\textwidth]{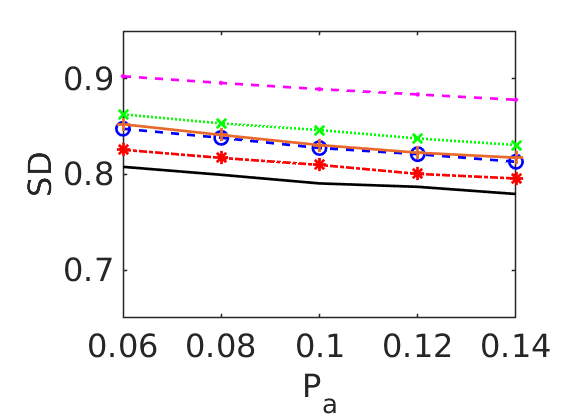}}
    \subfigure[Defense cost ($D_c$)]{
    \includegraphics[width=0.24\textwidth, height=0.19\textwidth]{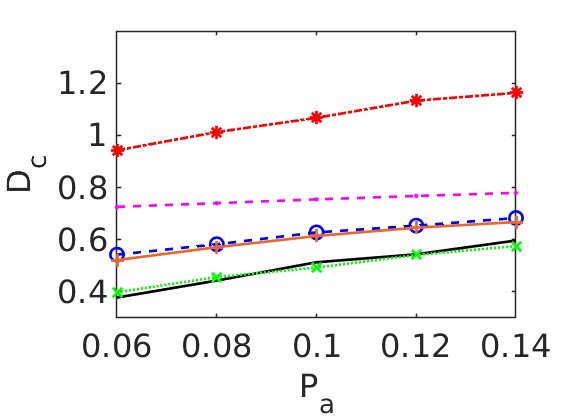}}\hspace{-0.85em}  
    \caption{Effect of varying the fraction of seeding attacks ($P_a$) under a random network.}
\label{fig_p_a_er}
\end{figure*}

Fig.~\ref{fig_p_a_er} demonstrates how the different levels of attack density impact the performance of all comparing schemes with respect to the performance metrics in Section 5.1 under the random network topology. The overall trends of the performance as more seeding attackers are added in the network are as follows. 
An increase of the fraction of seeding attackers decreases software diversity and the size of the giant component, while at the same time increasing the fraction of compromised nodes and defense costs. This latter effect is because more nodes become compromised and accordingly more site percolation-based adaptations are required by the IDS (i.e., disconnecting all edges to a detected, compromised node). 
Also across the range of the attack density and with respect to all metrics expect defense cost ($D_c$), the overall performance order clearly follows: SDA with $\rho = -0.6 \geq$ SDA with $\rho = 0 \geq$ SDA with $\rho = 1 \approx$ Random-A $\geq$ Random-Graph-C $\approx$ No-A. As for defense cost ($D_c$), the overall performance order follows: SDA with $\rho = 0 \approx$ No-A $\geq$ SDA with $\rho = 1 \approx$ Random-A $\geq$ SDA with $\rho = -0.6 \geq$ Random-Graph-C.

%fig4

%fig5
\begin{figure*}[!ht]
  \centering
  \subfigure{
    \includegraphics[width=0.65\textwidth, height=0.02\textwidth]{figs/fig2/legend.png}}\vspace{-1em}
   
   \setcounter{subfigure}{0}
  \subfigure[\% of compromised nodes ($P_c$)]{
    \includegraphics[width=0.24\textwidth, height=0.19\textwidth]{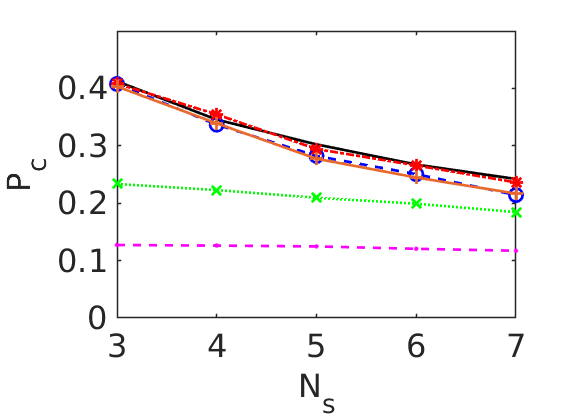}}\hspace{-0.85em}
  \subfigure[Size of the giant component ($S_g$)]{
    \includegraphics[width=0.24\textwidth, height=0.19\textwidth]{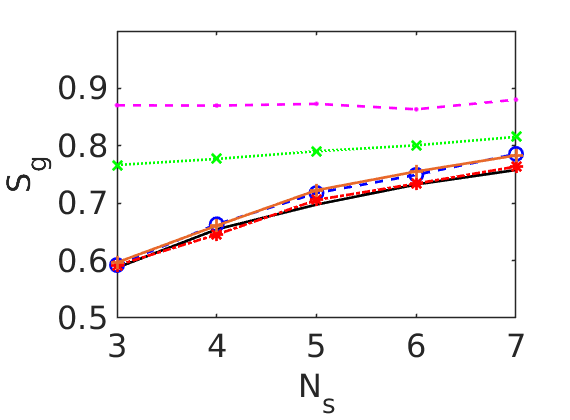}}
  \subfigure[Software diversity ($SD$)]{
    \includegraphics[width=0.24\textwidth, height=0.19\textwidth]{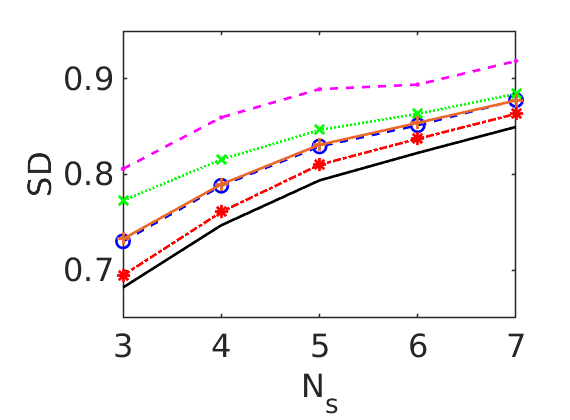}}
      \subfigure[Defense cost ($D_c$)]{
    \includegraphics[width=0.24\textwidth, height=0.19\textwidth]{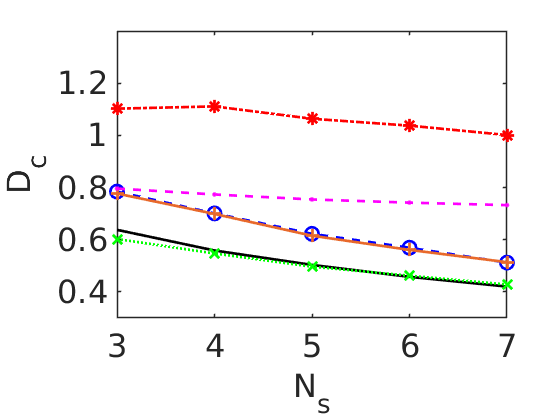}}\hspace{-0.85em}  
    \caption{Effect of the number of software packages $(N_s)$ under a random network.}
\label{fig_n_s_er}
\end{figure*}

\subsubsection{Effect of the Number of Software Packages ($N_s$) Under a Random Network} Fig.~\ref{fig_n_s_er} shows how a different number of software packages ($N_s$) available impacts the performance in terms of the metrics in Section 5.1 under the random network model. Similar to the previous results displayed in Figs.~\ref{fig_p_er} and \ref{fig_p_a_er}, the degree of software diversity is well aligned with that of the size of the giant component. But, noticeably, SDA with optimal $\rho = -0.6$ performs the best with high resiliency even under a very small $N_s$. In fact, this scheme shows steady performance in terms of all the metrics except software diversity ($SD$) across the range of $N_s$ considered in this work. This is due to the low network density and our intelligent edge adaptation. On the other hand, the other five schemes exhibit improved performance when $N_s$ is increased because the availability of more software packages can significantly contribute to increasing the degree of software diversity of each node. Unlike the previous results shown in Figs.~\ref{fig_p_er} (d) and \ref{fig_p_a_er} (d), Random-Graph-A attains maximum defense cost when $N_s=4$. This is due to two conflicting related factors: On the one hand, a node has more options to choose with more software packages available, i.e., a much lower chance to choose the same software package as the one the node currently has, thereby increasing the likelihood of shuffling and the shuffling cost. On the other hand, the network becomes more secure when more software packages are available, which results in lower IDS cost.

\section{Identification of Optimal $k$ under a Random Network}

%fig6
\begin{figure}[h]
  \centering
    \includegraphics[width=0.3\textwidth, height=0.22\textwidth]{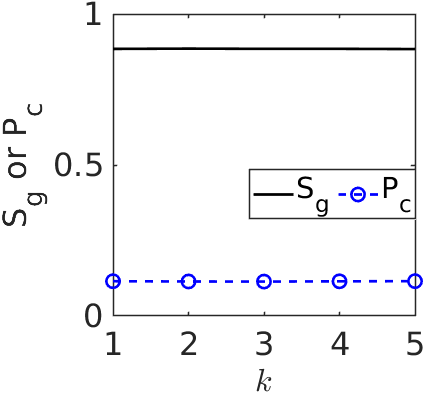}
    \caption{Effect of varying $k$ (a maximum hop distance) on network connectivity ($S_g$) and security vulnerability ($P_c$) under a random network.}
\label{fig_er_sensitivity_k}
\end{figure}

Fig.~\ref{fig_er_sensitivity_k} shows the sensitivity analysis when varying $k$ used in the software diversity metric of the SDA with $l=1$ and $\rho=-0.6$. We observe that both $S_g$ and $P_c$ are not sensitive when $k$ varies from $1$ to $5$. Thus, we use $k=1$ in our experiments to minimize computational complexity.

\section{Identification of Optimal $l$ under a Random Network}

%fig6
\begin{figure}[h]
  \centering
    \includegraphics[width=0.3\textwidth, height=0.22\textwidth]{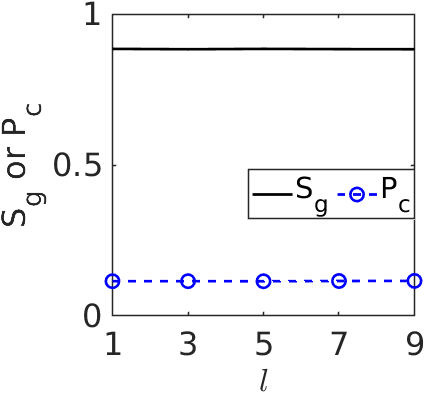}
    \caption{Effect of varying $l$ (a maximum number of attack paths) on network connectivity ($S_g$) and security vulnerability ($P_c$) under a random network.}
\label{fig_er_sensitivity_l}
\end{figure}
Fig.~\ref{fig_er_sensitivity_l} shows the sensitivity analysis under varying $l$ used in estimating the software diversity metric of the SDA when $k=1$ and $\rho=-0.6$.  We observe that both $S_g$ and $P_c$ are not sensitive to varying $l$ from $1$ to $9$. Thus, we set $l=1$ in our experiments to minimize computational complexity.
\bibliographystyle{plain}
\bibliography{diversity-bib}